\documentclass[12pt, draftclsnofoot, onecolumn]{IEEEtran}

\usepackage{amsmath,epsfig,amssymb,verbatim}
\usepackage{cite}
\usepackage{graphicx}
\usepackage{bm,bbm}
\usepackage{subfigure}

\newcommand{\E}{\mathbb{E}}

\newtheorem{Theorem}{Theorem}

\newtheorem{Corollary}{Corollary}

\begin{document}

\title{Age of Information in Downlink Systems: Broadcast or Unicast Transmission?}

\author{Zhifeng Tang,~\IEEEmembership{Student Member,~IEEE,}
Nan Yang,~\IEEEmembership{Senior Member,~IEEE,}\\
Parastoo Sadeghi,~\IEEEmembership{Senior Member,~IEEE,}
and Xiangyun Zhou,~\IEEEmembership{Senior Member,~IEEE}\vspace{-1em}

\thanks{This work was supported by the Australian Research Council Discovery Project (DP180104062).}
\thanks{Z. Tang, N. Yang, and X. Zhou are with the School of Engineering, Australian National University, Canberra, ACT 2600, Australia (Email: \{zhifeng.tang, nan.yang, xiangyun.zhou\}@anu.edu.au).}
\thanks{P. Sadeghi is with the School of Engineering and Information Technology, University of New South Wales, Canberra, ACT 2612, Australia (Email: p.sadeghi@unsw.edu.au).}}


\maketitle

\begin{abstract}
We analytically decide whether the broadcast transmission scheme or the unicast transmission scheme achieves the optimal age of information (AoI) performance of a multiuser system where a base station (BS) generates and transmits status updates to multiple user equipments (UEs). In the broadcast transmission scheme, the status update for all UEs is jointly encoded into a packet for transmission, while in the unicast transmission scheme, the status update for each UE is encoded individually and transmitted by following the round robin policy. For both transmission schemes, we examine three packet management strategies, namely the non-preemption strategy, the preemption in buffer strategy, and the preemption in serving strategy. We first derive new closed-form expressions for the average AoI achieved by two transmission schemes with three packet management strategies. Based on them, we compare the AoI performance of two transmission schemes in two systems, namely, the remote control system and the dynamic system. Aided by simulation results, we verify our analysis and investigate the impact of system parameters on the average AoI. For example, the unicast transmission scheme is more appropriate for the system with a large number UEs. Otherwise, the broadcast transmission scheme is more appropriate.
\end{abstract}

\begin{IEEEkeywords}
Age of information, multiuser system, short packet communications, packet management.
\end{IEEEkeywords}

\IEEEpeerreviewmaketitle

\section{Introduction}\label{sec:introduction}

Real-time applications, such as intelligent transport systems and factory automation, have recently attracted rapidly increasing interests from academia and industry. In these applications, timely status update plays an indispensable role in accurate monitoring and control \cite{Simsek2016,Li2019,Hou2021Mag,She2021Pro}. To reduce transmission latency in real-time applications, short packet communication has been widely considered as a promising solution, due to its unique benefits in delay reduction \cite{Huang2019,yuan2021performance,Sun20181,Chunhui2022}. Moreover, in order to fully characterize the freshness of delivered status information, the concept of age of information (AoI) has been introduced as a new and effective performance metric \cite{Kaul2011}. Specifically, the AoI is defined as the elapsed time since the last successfully received status update being generated by the transmitter, which is a time metric capturing both latency and freshness of transmitted status information.

Since being introduced in \cite{Kaul2011}, the concept of AoI has reaped a wide range of attention and interests. The authors of \cite{Kaul2012} studied the average AoI in a first-come-first-served (FCFS) single-user system. Different from \cite{Kaul2012}, \cite{Inoue2019} proposed the last-come-first-served (LCFS) queuing policy, which was shown to achieve a lower average AoI than the FCFS queuing policy. Building upon these efforts on the single-user system, increasing research efforts have been devoted to investigating the AoI performance of multiuser systems. The authors of \cite{Yates2012} extended \cite{Kaul2012} to analyze the average AoI under the FCFS queuing policy in a multiuser system. By considering the effect of unreliable channels on packet loss, \cite{Yates2017} introduced a feedback mechanism to deliver generated packets as timely as possible. Moreover, \cite{Jiang2019} examined the average AoI for three different scheduling policies, i.e., round robin (RR), work-conserving non-collision, and random access. The authors of \cite{Jiang2019} further pointed out that the RR policy is the optimal arrival-independent scheduling policy to minimize the average AoI. By considering sporadic packet generation rates of users, \cite{Tang2020} proposed a random access based transmission scheme to improve the average AoI performance. In addition, \cite{Tang2022Lt} designed a Whittle index based scheduling policy to optimize the AoI performance while considering the effect of unreliable channels.

To improve the AoI performance, multiple packet management strategies were discussed, e.g., \cite{Kaul2012b,Yates2019,Chen2016Presence,Arafa2019HARQ,Tang2021PM}. As an early study, \cite{Kaul2012b} introduced two different packet management strategies for the LCFS system, namely, the preemption strategy and the non-preemption strategy in a single-user system. In the preemption strategy, when a new packet is generated at the source, it is allowed to replace the current packet in service. In contrast, for the non-preemption strategy, the newly generated packet has to wait for the current packet in service to be transmitted. The authors of \cite{Yates2019} extended \cite{Kaul2012b} into a multiuser system and revealed the superiority of the preemption strategy when the packet generation rate is large. A retransmission strategy was proposed in \cite{Chen2016Presence}, where the source keeps transmitting the most recent packet once the transmission is completed. \cite{Chen2016Presence} further pointed out that the retransmission strategy significantly reduces the peak AoI (PAoI) when the transmission error rate is high. In \cite{Arafa2019HARQ}, a hybrid automatic repeat request (HARQ) protocol was employed such that several incremental redundancy bits are sent when the receiver cannot successfully decode. Very recently, \cite{Tang2021PM} introduced a packet dropping strategy and demonstrated the benefit of this strategy on decreasing the average AoI.

Motivated by the benefits of short packet communications on latency reduction, the AoI performance of short packet communications was analyzed to evaluate the impact of short packets on the freshness of transmitted information, e.g., \cite{devassy2018delay,Devassy2019,Basnay2021,WangGC2019,Sung2022wcl,Yu2020tvt}. Specifically, \cite{devassy2018delay} investigated the impact of the packet blocklength on the delay and the PAoI in a single-user system. Considering the same system, \cite{Devassy2019} extended \cite{devassy2018delay} to analyze the probability of the peak-age violation exceeding a threshold. Focusing on a decode-and-forward relaying system, \cite{Basnay2021} estimated the impact of the packet generation rate, the packet blocklength, and the blocklength allocation factor on the average AoI. Moreover, \cite{WangGC2019} studied the optimal packet blocklength of non-preemption and preemption strategies for minimizing the average AoI, while \cite{Sung2022wcl} derived the average AoI and the AoI violation probability in a downlink system. Considering a feedback mechanism, \cite{Yu2020tvt} derived the average AoI under two protocols, i.e., the traditional protocol and the ARQ protocol, and presented sub-optimal blocklengths to minimize the average AoI for both protocols.

Although the aforementioned studies have investigated the impact of the packet blocklength on the AoI performance of short packet communication systems, the impact of different packet management strategies on the AoI performance of multiuser short packet communication systems has not been touched. Moreover, the impact of correlation among the information for user equipments (UEs) on the average AoI has not been studied in the literature. To the best of our knowledge, this is the \textit{first} work to investigate transmission scheme selection to optimize the AoI performance based on the correlation among the information for UEs. The main contributions of this paper are summarized as follows:
\begin{itemize}
   \item We derive new closed-form expressions for the average AoI of a multiuser system, where we consider two transmission schemes with three packet management strategies, i.e., the non-preemption strategy, the preemption in buffer strategy, and the preemption in serving strategy. Aided by simulations, we demonstrate the accuracy of our analytical results. We also find that the block error rate has a more significant impact on the average AoI for the unicast transmission scheme than the broadcast transmission scheme. We further find that the non-preemption strategy achieves a lower average AoI than the preemption in serving strategy in the broadcast transmission scheme, and the preemption in buffer strategy achieves the lowest average AoI compared with the non-preemption and the preemption strategy in serving strategies in the unicast transmission scheme.
    \item Considering the information correlation among UEs, we compare the average AoI achieved by both transmission schemes with non-preemption strategy in a remote control system. In this system, with the feature of stochastic status update generation, we derive the threshold of the information ratio between the broadcast transmission scheme and the unicast transmission scheme to decide which transmission scheme is adopted. We then show the relationship between this threshold and the number of UEs in two special cases. We further find that the unicast transmission scheme achieves a lower average AoI than the broadcast transmission scheme when the number of UEs is large.
    \item 
    We then extend the comparative study of the two transmission schemes to a dynamical system. In this system, we approximate the expected average AoI under the zero-waiting policy. Based on the approximation result, we determine the approximated threshold of the ratio between the individual information and the common information to decide which transmission scheme is adopted. We observe that the increase in this ratio has a more pronounced impact on the average AoI for the broadcast transmission scheme than the unicast transmission scheme.
\end{itemize}

The rest of the paper is organized as follows. In Section \ref{Sec:System}, the system model and two transmission schemes are described. New closed-form expressions for the average AoI of two transmission schemes with different packet management strategies are derived in Section \ref{Sec:Derivation}. In Section \ref{Sec:Transselect}, the transmission schemes are decided based on the AoI performance in two systems. The numerical results are discussed in Section \ref{Sec:Numerical} and the paper is concluded in Section \ref{Sec:Conclusion}.

\section{System Model}\label{Sec:System}

\subsection{System Description}

\begin{figure}[!t]
    \centering
    \includegraphics[width=0.85\columnwidth]{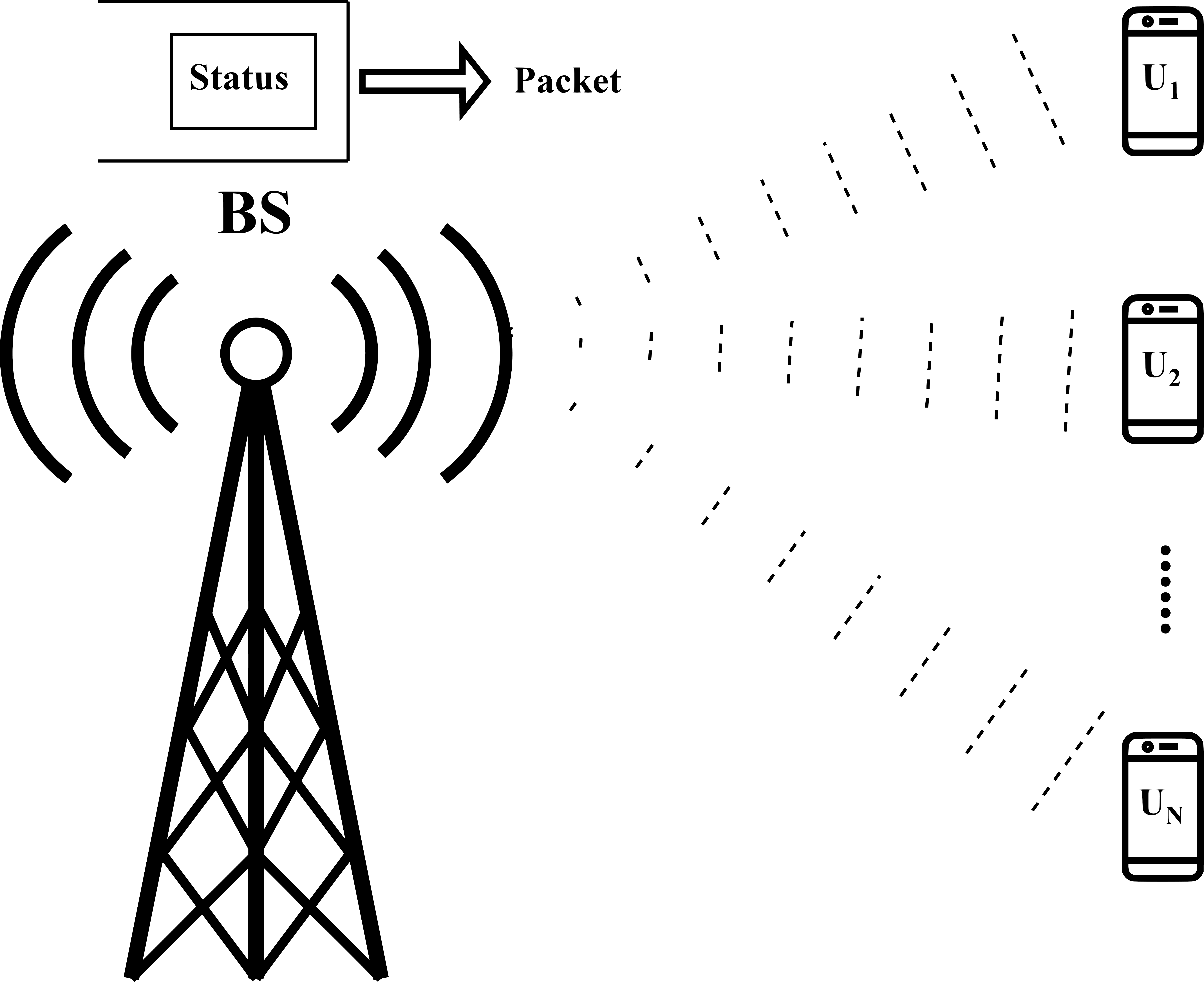}
    \caption{Illustration of our considered multiuser system where the BS transmits status update to $N$ UEs.}\label{fig:system_model}
    \vspace{-1em}
\end{figure}

We consider a multiuser wireless communication system, as depicted in Fig.~\ref{fig:system_model}, where a base station (BS) transmits status update to $N$ UEs. We denote the $n$th UE by $U_{n}$, where $n=1,2,\cdots,N$. In this system, the BS generates the status update for $N$ UEs according to a Poisson process with the rate $\lambda$. Then, packets are generated by the BS based on the generated status update, and the packets are transmitted from the BS to UEs. In this system, we consider two transmission schemes, namely, the broadcast transmission scheme and the unicast transmission scheme, described as follows:
\begin{itemize}
    \item In the broadcast transmission scheme, we assume that the status update for $N$ UEs contains $L$ bits, where $L$ is a fixed value. Once the BS generates the status update, it encodes these $L$ bits into a packet with the blocklength of $M$ channel use (c.u.) and transmits this packet to $N$ UEs. For this transmission, we define the coding rate, $R$, as the ratio between the number of bits in the status update and the blocklength of the transmitted packet, i.e., $R=\frac{L}{M}$.
    \item In the unicast transmission scheme, the BS separately transmits the status update to $N$ UEs by following an RR policy. We denote the length of the status update and the packet blocklength for $U_n$ by $L_n$ and $M_n$, respectively. For this transmission, we define the coding rate for $U_n$, $R_n$, as the ratio between the number of bits in the status update and the blocklength of the transmitted packet, i.e., $R_n=\frac{L_n}{M_n}$.
\end{itemize}

We clarify that the total number of bits of the status update for $N$ UEs is $L$ in the broadcast transmission scheme and $\sum_{n=1}^{N}L_n$ in the unicast transmission scheme. According to the information correlation among UEs, 
the joint information is less than or equal to the sum of individual information, i.e., $L\leq \sum_{n=1}^{N}L_n$. Thus, we denote $\alpha$ as the information ratio between the broadcast transmission scheme and the unicast transmission scheme and express it by $\alpha = \frac{L}{\sum_{n=1}^{N}L_n}$, where $0<\alpha\leq1$.

We assume an additive white Gaussian noise (AWGN) channel between the BS and each UE, which has been widely considered in the literature, e.g., \cite{WangGC2019}. We clarify that the results obtained for the AWGN channel can be easily extended to other channels. According to \cite{Polyanskiy2010}, the block error rate for $U_n$ using finite block length coding can be approximated as
\begin{align}\label{eq:blockerrrate}
    \epsilon_n(l,m,\gamma_n) = Q\left(\frac{\frac{1}{2}\log_2(1+\gamma_n)-\frac{l}{m}}{\log_2(e)\sqrt{\frac{1}{2m}\left(1-\frac{1}{(1+\gamma_n^2)}\right)}}\right),
\end{align}
where $l$ is the number of bits in the status update, $m$ is the packet blocklength, $\gamma_n$ is the received signal-to-noise ratio (SNR) at $U_n$, and $Q(x)=\int_{x}^{\infty}\!\frac{1}{\sqrt{2\pi}}\!\exp\left(\!-\frac{t^2}{2}\!\right)\!\mathrm{d}t$ is the $Q$-function. We note that \eqref{eq:blockerrrate} is very tight for short packet communications when $m\!\geq\!100$ \cite{Polyanskiy2010}. For 
our considered system, we denote $T_u$ as the transmission time for each c.u.. To facilitate our analysis, we define $T_u$ as the unit time.

We note that, apart from the transmission time, the BS requires an extra pre-processing and status update processing time to send connection requests for establishing the transmission link with each UE and processing the status update for each UE in the unicast transmission scheme, comparing with the broadcast transmission scheme \cite{Ji2018URLLC}. Here, we denote $M_{\textrm{L}}$ as the time spent on the pre-processing and status update processing for each UE. Hence, we define the serving time as $M_{\textrm{L}}+M_n$ and the ratio between the transmission time and the serving time for $U_n$ is $\rho_n = \frac{M_n}{M_{\textrm{L}}+M_n}$ in the unicast transmission scheme.

\subsection{Packet Management Strategies}

To ensure the freshness of packets, we assume that the LCFS queuing policy is adopted and the buffer at the BS only stores the newest status update. 
In the broadcast transmission scheme, when a new status update is generated and the BS is in the idle state, the BS encodes the status update into a packet and transmits this packet to all UEs. When a new status update is generated at the BS and the BS is transmitting a packet, we examine two packet management strategies, namely, the non-preemption strategy and the preemption in serving strategy, detailed as follows:
\begin{itemize}
    \item Broadcast transmission with non-preemption (BRNP) strategy: In this strategy, if a new status update is generated at the BS and the BS is transmitting a packet, the BS stores the new status update in the buffer and keeps the current transmission.
    \item Broadcast transmission with preemption in serving (BRPS) strategy: In this strategy, preemption is considered such that the newly generated status update always preempts the transmission of the current status update. Specifically, if a new status update is generated at the BS and the BS is transmitting a status update, the BS discards the current transmission and starts to transmit the new status update.
\end{itemize}

\begin{figure}[t]
    \centering
\subfigure[DNP strategy.]{
    \includegraphics[width=0.95\columnwidth]{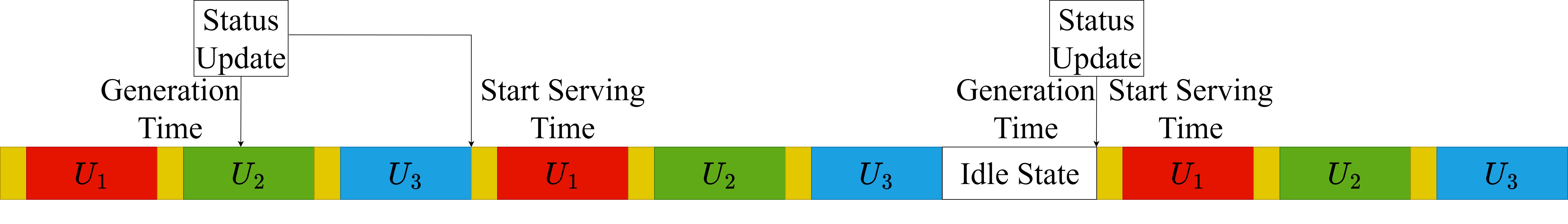}\label{fig:DNP}
    }
\vspace{-0.5em}
\subfigure[DPB strategy.]{
    \includegraphics[width=0.95\columnwidth]{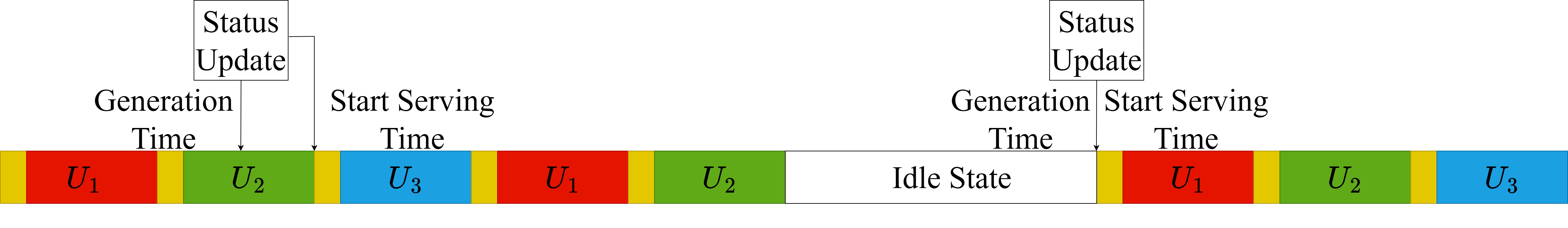}\label{fig:DPB}}
 \vspace{-0.5em}   
\subfigure[DPS strategy.]{
    \includegraphics[width=0.95\columnwidth]{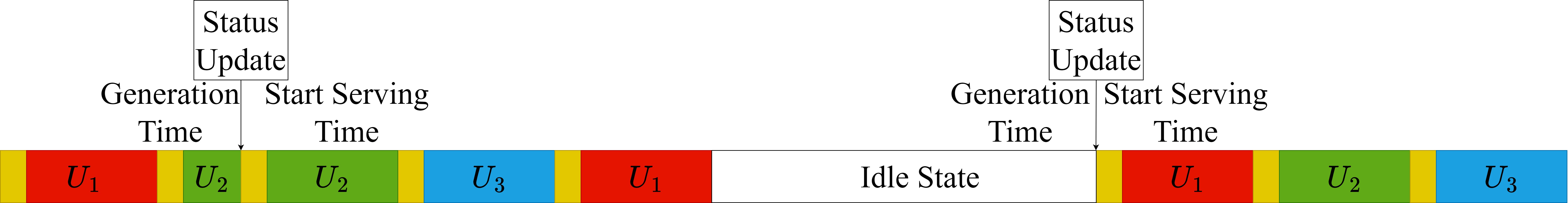}\label{fig:DPs}}
    \centering
    \caption{An example to illustrate the timeliness of the unicast transmission scheme with three packet management strategies in a three-UE system.}\label{fig:DistributedSt}

    \vspace{-1.5em}
\end{figure}

In the unicast transmission scheme, when a new status update is generated and the BS is in the idle state, the BS processes the status update and transmits the packets to $N$ UEs, i.e., $U_1$, $U_2$, $\cdots$, $U_N$. In particular, the BS processes the status update and encodes the required status update information for $U_n$ into a packet. Then the BS establishes the transmission link with $U_n$ and transmits this packet. After the transmission of $U_n$'s packet, the BS repeats this step for $U_{n+1}$ until all UEs are served. We note that the BS sends the status update only once for each UE. When a new status update is generated at the BS and the BS is serving a packet, we examine three packet management strategies, namely the non-preemption strategy, the preemption in buffer strategy, and the preemption in serving strategy, detailed as follows:
\begin{itemize}
\item Unicast transmission with non-preemption (DNP) strategy: In this strategy, if a new status update is generated at the BS and the BS is serving a UE, the BS stores the new status update in the buffer and keeps current service. We note that the serving order of each status update is always from $U_1$ to $U_N$ in this strategy.
\item Unicast transmission with preemption in buffer (DPB) strategy: In this strategy, when a new status update is generated at the BS and the BS is serving $U_n$, the BS stores the new status update in the buffer. After the transmission of $U_n$'s packet, the BS processes this status update and transmits it to $U_{n+1}$.
\item Unicast transmission with preemption in serving (DPS) strategy: In this strategy, preemption is considered such that the newly generated status update always preempts the service of the current status update. Specifically, if a new status update is generated at the BS and the BS is serving $U_n$, the BS discards the current serving and starts to serve the new status update for $U_n$.
\end{itemize}
In Fig~\ref{fig:DistributedSt}, we provide an example of the timeliness of the Unicast transmission scheme with three packet management strategies in a three-UE system. We note that the BRNP strategy and the BRPS strategy can be considered as the special case of the DNP strategy and the DPS strategy with $N=1$ and $M_{\textrm{L}}=0$, respectively.

\subsection{Formulation of AoI}

\begin{figure}[t]
    \centering
    \includegraphics[width=0.8\columnwidth]{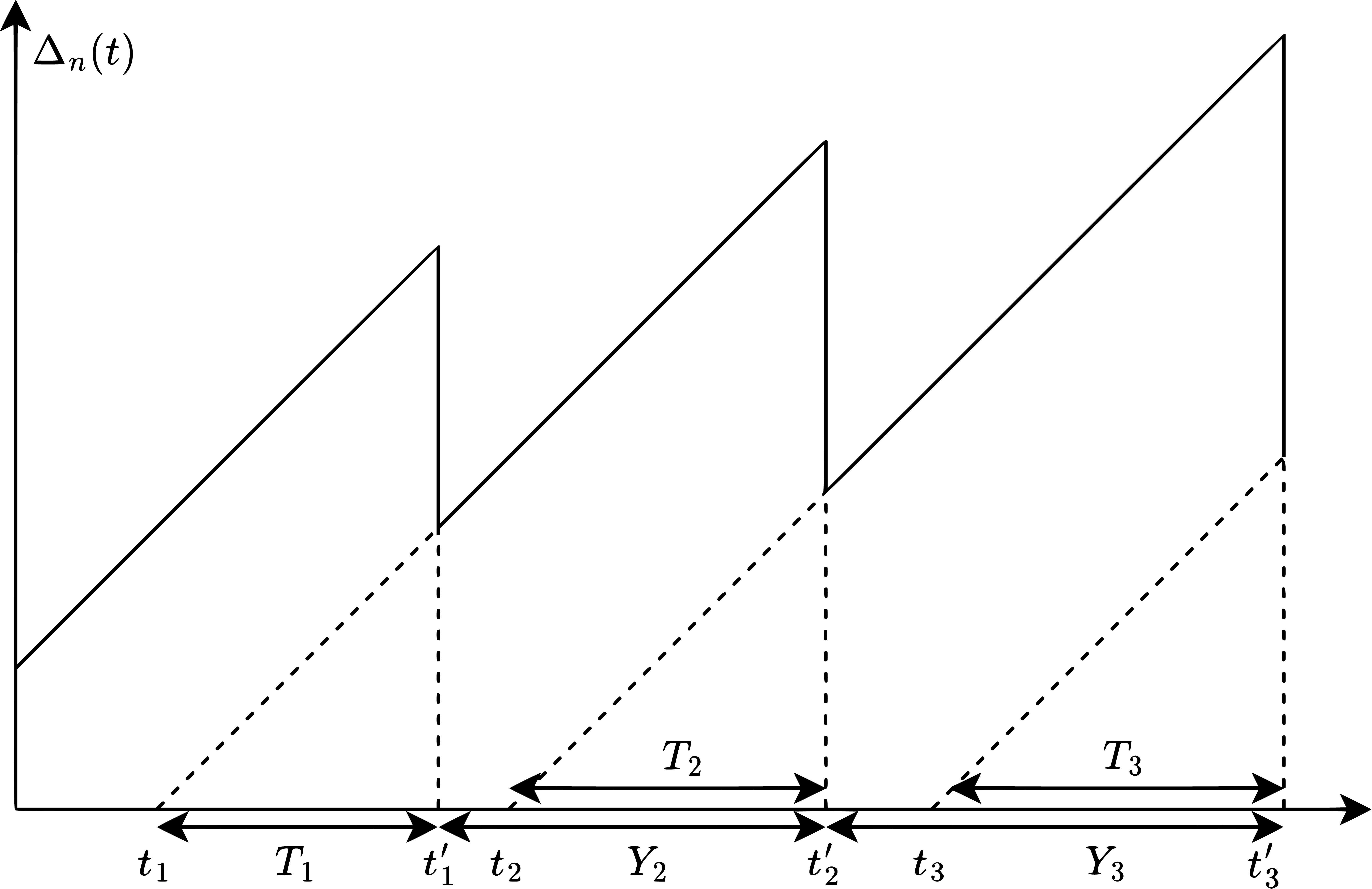}
    \centering
    \caption{AoI variation of the selected UE, $U_n$.}\label{fig:AoIevo}
\end{figure}

In this subsection, we formulate the average AoI of the considered system. Without loss of generality, we arbitrarily select one UE, $U_n$, and analyze its average AoI, $\Delta_n$. We denote $\Delta_n(t)$ as the AoI of $U_n$ at time slot $t$. 
Fig.~\ref{fig:AoIevo} plots a sample variation $\Delta_n(t)$ as a function of $t$. We assume that the observation begins at $t=0$ with the AoI of $\Delta_n(0)$. From Fig.~\ref{fig:AoIevo}, we express the AoI at time $t$ as
\begin{align}\label{eq:2}
    \Delta_n(t) = t - u_n(t),
\end{align}
where $u_n(t)$ is the generation time of the most recently received status update at $U_n$ at time $t$. Then, the time-average AoI of $U_n$ over the observation time interval $(0,\tau)$ is calculated as
\begin{align}\label{eq:AoIEQ5}
    \Delta_n = \frac{1}{\tau}\int_{0}^{\tau}\Delta_n(t) \mathrm{d}t.
\end{align}

We denote $P_{j}$ as the $j$th successfully received status update generated for $U_n$ after time $t=0$, $j=1,2,\cdots$. We then denote $Y_j$ as the time interval from the received time of $P_{j-1}$ to the received time of $P_{j}$ and denote $T_j$ as the time interval from the generation time of $P_{j}$ to the received time of ${P}_{j}$. Therefore, we express $Y_{j}$ and $T_{j}$ as
\begin{align}
Y_j = t_{j}'-t_{j-1}'
\end{align}
and
\begin{align}
T_j = t_j'-t_j,
\end{align}
respectively, where $t_j$ is the generation time of $P_{j}$ and $t_j'$ is the time for $U_n$ to receive ${P}_{j}$. According to \cite{Kaul2012b}, the average AoI of $U_n$ is thus calculated as
\begin{align}\label{eq:newAoIevo}
\Delta_n= \frac{\E[Y_j^2]}{2\E[Y_j]}+\E[T_{j}],
\end{align}
where $\E[\cdot]$ denotes the expectation. By averaging $\Delta_n$ over all UEs, we obtain the average AoI of the considered system as
\begin{align}\label{eq:AveAoIallUE}
    \Delta = \frac{1}{N} \sum_{n=1}^{N}\Delta_n.
\end{align}

\section{Closed-form Analysis of Average AoI}\label{Sec:Derivation}

In this section, we derive new closed-form expressions for the average AoI of $U_n$ in two transmission schemes with different packet management strategies. We first focus on the DNP strategy and derive the closed-form expression for its average AoI in the following Theorem.

\begin{Theorem}\label{Theorem:1}
In the DNP strategy, the closed-form expression for the average AoI of $U_n$ with the received SNR, $\gamma_n$, is derived as
\begin{align}\label{eq:expreAoIpar}
\Delta_{\mathrm{DNP},n} =& \frac{1+\epsilon_n}{2(1-\epsilon_n)}\left(M_{\textrm{T}}+\frac{1}{\lambda}e^{-\lambda M_{\textrm{T}}}\right)+\frac{(2e^{-\lambda M_{\textrm{T}}}-e^{-2\lambda M_{\textrm{T}}})}{2(\lambda^2M_{\textrm{T}}+\lambda e^{-\lambda M_{\textrm{T}}})}\notag\\
&-\sum_{k=n+1}^N M_k'+\left(\frac{1}{\lambda}+M_{\textrm{T}}\right)(1-e^{-\lambda M_{\textrm{T}}}),
\end{align}
where $M_{k}' = M_k + M_{\textrm{L}}$, $M_{\textrm{T}} = \sum_{k=1}^N M_k'$, and $\epsilon_n=\epsilon(L_n,M_n,\gamma_n)$.
\begin{IEEEproof}
See Appendix \ref{Appendix:A}.
\end{IEEEproof}
\end{Theorem}

Using Theorem~\ref{Theorem:1}, the average AoI of the system in the DNP strategy can be obtained by averaging $\Delta_n$ over all UEs. We find that the average AoI of the system is minimized if $M_n$ is monotonically non-decreasing with respect to $n$, i.e., $M_{n_1}\geq M_{n_2}$, for any $n_1>n_2$. It indicates that the UE with the shortest blocklength needs to be served first in the DNP strategy.

We next derive the average AoI in the DPB strategy. To facilitate our derivation, we define $k_n = k+n$ if $k+n\leq N$, but $k_n=k+n-N$ if $k+n>N$, and present the closed-form expression for the AoI in the following theorem. 
\begin{Theorem}\label{Theorem:2}
In the DPB strategy, the closed-form expression for the average AoI of $U_n$ with the received SNR, $\gamma_n$, is derived as
\begin{align}\label{eq:expreAoIS2}
\Delta_{\mathrm{DPB},n}=& \frac{1+\epsilon_n}{2(1-\epsilon_n)}\left(M_{\textrm{T}}+\xi\right)
+\frac{1-e^{-\lambda M_{\textrm{T}}}}{\lambda} + \frac{2\xi-\lambda \xi^2}{2\lambda(M_{\textrm{T}}+\xi)}\notag\\
&-\sum_{k=1}^{N}M_{k_n}'\left(\frac{e^{-\lambda M_{\textrm{T}}}\left(1-e^{-\lambda\sum_{\kappa=k}^{N-1}M_{\kappa_n}'}\right)}
{1-e^{-\lambda M_{\textrm{T}}}}\right)+M_n'(1-e^{-\lambda M_{\textrm{T}}}),
\end{align}
where $\xi = \frac{e^{-\lambda M_{\textrm{T}}}}{\lambda(1-e^{-\lambda M_{\textrm{T}}})}\sum_{k=1}^{N}(1-e^{-\lambda M_{k_n}'})$.
\begin{IEEEproof}
See Appendix \ref{Appendix:B}.
\end{IEEEproof}
\end{Theorem}

To improve the freshness of information, the system is always designed to eliminate the waiting time in the buffer, which is the zero-waiting policy. Under this policy, we obtain the closed-form expressions for the average AoI in the DNP and DPB strategies in the following Corollary.
\begin{Corollary}\label{Corollary:1}
With the received SNR, $\gamma_n$, the closed-form expression for the average AoI of $U_n$ in the DNP strategy under the zero-waiting policy is given by
\begin{align}\label{eq:DNPZzwp}
\Delta_{\mathrm{DNPZ},n} = \frac{1+\epsilon_n}{2(1-\epsilon_n)}M_{\textrm{T}}+M_{\textrm{T}}-\sum_{k=n+1}^N M_k',
\end{align}
and the closed-form expression for the average AoI of $U_n$ in the DPB strategy under the zero-waiting policy is given by
\begin{align}\label{eq:DPBZzwp}
\Delta_{\mathrm{DPBZ},n} = \frac{1+\epsilon_n}{2(1-\epsilon_n)}M_{\textrm{T}}+M_n'.
\end{align}
\begin{IEEEproof}
Under the zero-waiting policy, we obtain $\Delta_{\mathrm{DNPZ},n} =\lim_{\lambda\rightarrow \infty}\Delta_{\mathrm{DNP},n}$ and $\Delta_{\mathrm{DPBZ},n} =\lim_{\lambda\rightarrow \infty}\Delta_{\mathrm{DPB},n}$, resulting in \eqref{eq:DNPZzwp} and \eqref{eq:DPBZzwp}.
\end{IEEEproof}
\end{Corollary}

From Corollary~\ref{Corollary:1}, we find that the DPB strategy always outperforms the DNP strategy by achieving a lower average AoI under the zero-waiting policy. This is due to the fact that comparing with the DNP strategy, the DPB strategy eliminates the waiting time of status update for each UE caused by the service of other UEs.

We further derive the average AoI in the DPS strategy in the following Theorem.
\begin{Theorem}\label{Theorem:3}
In the DPS strategy, the closed-form expression for the average AoI of $U_n$ with the received SNR, $\gamma_n$, is derived as \eqref{eq:expreAoIS3}, where $p_k=1-e^{-\lambda M_{k_n}'}$ and $\psi = \frac{e^{-\lambda M_{\textrm{T}}}}{1-e^{-\lambda M_{\textrm{T}}}}\sum_{k=1}^{N}\frac{p_k}{1-p_k}$.
\begin{IEEEproof}
See Appendix \ref{Appendix:C}.
\end{IEEEproof}
\end{Theorem}
\begin{figure*}[!t]
\begin{align}\label{eq:expreAoIS3}
\Delta_{\mathrm{DPS},n}&=\frac{1+\epsilon_n}{2\lambda(1-e^{-\lambda M_{\textrm{T}}})(1-\epsilon_n)}\sum_{k=1}^{N}\frac{p_k}{1-p_k}+\sum_{k=1}^{N}M_{k_n}'\frac{e^{-\lambda\sum_{\kappa=k+1}^{N}M_{\kappa_n}'}-e^{-\lambda M_{\textrm{T}}}}{1-e^{-\lambda M_{\textrm{T}}}}\notag\\
&+\frac{\lambda e^{-\lambda M_{\textrm{T}}}}{2\psi}\left[\frac{2\psi-\psi^2}{\lambda^2}+\sum_{k=1}^{N}\left(\frac{p_k}{1-p_k}\left(\frac{2}{\lambda^2}
+\frac{p_k}{\lambda^2(1-p_k)}\right)-\frac{2M_{k_n}'}{\lambda(1-p_k)}\right)\right.\notag\\
&\left.-\frac{2e^{-\lambda M_{\textrm{T}}}}{1-e^{-\lambda M_{\textrm{T}}}}\sum_{k=1}^{N}\frac{p_k}{1-p_k}\sum_{\kappa=1}^{k-1}
\frac{p_{\kappa}}{1-p_{\kappa}}\left(\frac{1}{\lambda}-\frac{M_{\kappa_n}'(1-p_\kappa)}
{p_\kappa}\right)\right].
\end{align}
\hrulefill
\vspace{-1em}
\end{figure*}

Since the BRNP strategy and the BRPS strategy are the special case of the DNP strategy and the DPS strategy with $N=1$ and $M_{\textrm{L}}=0$, respectively, we obtain the closed-form expressions for the average AoI in the BRNP strategy and the BRPS strategy in the following Corollary.
\begin{Corollary}\label{Corollary:2}
The closed-form expressions for the average AoI of $U_n$ with the received SNR, $\gamma_n$, in the BRNP strategy and the BRPS strategy are given by
\begin{align}\label{eq:expreAoIB1}
\Delta_{\mathrm{BRNP},n}=& \frac{1+\epsilon_n}{2(1-\epsilon_n)}\left(M+\frac{1}{\lambda}e^{-\lambda M}\right)+\frac{2e^{-\lambda M}-e^{-2\lambda M}}{2\left(\lambda^2M+\lambda e^{-\lambda M}\right)}+\left(\frac{1}{\lambda}+M\right)\left(1-e^{-\lambda M}\right)
\end{align}
and
\begin{align}\label{eq:expreAoIB2}
\Delta_{\mathrm{BRPS},n}=\frac{1}{\lambda e^{-\lambda M}(1-\epsilon_n)},
\end{align}
respectively.
\begin{IEEEproof}
By substituting $L=L_1$, $M=M_1$, $N=1$, and $M_{\textrm{L}}=0$ into \eqref{eq:expreAoIpar} and \eqref{eq:expreAoIS3}, we obtain the closed-form expressions for the average AoI of $U_n$ in the BRNP strategy and the BRPS strategy, given by \eqref{eq:expreAoIB1} and \eqref{eq:expreAoIB2}, respectively.
\end{IEEEproof}
\end{Corollary}

From Corollary~\ref{Corollary:2}, we find that the average AoI monotonically decreases with respect to the status update generation rate, $\lambda$, in the BRNP strategy. Moreover, we find that the optimal status update generation rate is $\lambda=\frac{1}{M}$ in the BRPS strategy.

\section{Transmission Scheme Selection}\label{Sec:Transselect}

In this section, we investigate the selection of appropriate transmission for two systems, namely, a remote control system and a dynamic system, which correspond to a remote control automation factory and a vehicular network, respectively. To facilitate this investigation, we compare the average AoI achieved by both transmission schemes with the non-preemption strategy, i.e., the BRNP strategy and the DNP strategy, against the information ratio, $\alpha$. Specifically, we assume that all UEs require the same length of the status update, i.e., $L_n=L_h$, $n=1,2,\cdots,N$.

\subsection{Remote Control System}

In a remote control automation factory, a large number UEs are employed and the distances between the control BS and these UEs are almost same. Moreover, these UEs may require to cooperate with each other \cite{Liang2019}, implying that the required information between the cooperated UEs is highly correlated. To characterize these properties in a remote control automation factory, we consider a remote control system in this subsection and compare the average AoI achieved by the BRNP strategy and the DNP strategy against the information ratio, $\alpha$. In this system, we assume that UEs have the same received SNR, i.e., $\gamma_n=\gamma_h$, and consider the same coding rate, $R$, in the two strategies, i.e., $R=\frac{L}{M}=\frac{L_h}{M_h}$, where $M_n=M_h$ and $\rho=\rho_n$, $n = 1,2,\cdots,N$. We also assume that the coding rate is low, leading to a low block error rate, which is due to the fact that high reliable communication scenarios are of great interests in practice. Theorem~\ref{Lemma:1} characterizes the condition of $\alpha$ where the average AoI in one strategy is better than another.
\begin{Theorem}\label{Lemma:1}
In the remote control system, the BRNP strategy has a better AoI performance than the DNP strategy, if $\alpha\leq\alpha_{\textrm{th}}$, and the DNP strategy has a better AoI performance than the BRNP strategy, if $\alpha>\alpha_{\textrm{th}}$. Specifically, the information ratio threshold $\alpha_{\textrm{th}}$ satisfies
\begin{align}\label{eq:Threshold1}
&\left(\frac{3}{2}-e^{-\lambda\alpha_{\textrm{th}} \rho M_{\textrm{T}}}\right)\alpha_{\textrm{th}} \rho M_{\textrm{T}}+\frac{1}{2\lambda}\Omega\left(\lambda\alpha_{\textrm{th}} \rho M_{\textrm{T}}\right)=\left(\frac{2N+1}{2N}-e^{-\lambda M_{\textrm{T}}}\right) M_{\textrm{T}}
+\frac{1}{2\lambda}\Omega\left(\lambda M_{\textrm{T}}\right),
\end{align}
where $\Omega(\omega)=\frac{2e^{-\omega}-e^{-2\omega}}{\omega+e^{-\omega}}-e^{-\omega}$.
\begin{IEEEproof}
See Appendix~\ref{Appendix:D}.
\end{IEEEproof}
\end{Theorem}

Theorem~\ref{Lemma:1} reveals that the BRNP strategy is better when there is high correlation among the information for UEs, but the DNP strategy is better when there is low correlation among the information for UEs. Moreover, we find that the threshold $\alpha_{\textrm{th}}$ monotonically decreases as the ratio between the transmission time and the serving time, $\rho$, increases. This indicates that the BRNP strategy has a better AoI performance than the DNP strategy when the time spent on the pre-processing and status update processing, $M_{\textrm{L}}$, is long.

We then examine the value of $\alpha_{\textrm{th}}$ under the zero-waiting policy, i.e., $\lambda\rightarrow \infty$, and the sporadic status update generation rate, i.e., $\lambda\rightarrow 0$, in the following Corollary.
\begin{Corollary}\label{Corollary:3}
The information ratio threshold $\alpha_{\textrm{th}}$ under the zero-waiting policy is given by
\begin{align}\label{eq:thres1linf}
\alpha_{\textrm{th}} = \frac{2N+1}{3N\rho},
\end{align}
and $\alpha_{\textrm{th}}$ under the sporadic status update generation rate is given by
\begin{align}\label{eq:thres1l0}
\alpha_{\textrm{th}} = \frac{N+1}{2N\rho}.
\end{align}
\begin{IEEEproof}
When $\lambda\rightarrow \infty$, \eqref{eq:Threshold1} can be rewritten as
\begin{align}
\frac{3}{2}\alpha_{\textrm{th}}\rho M_{\textrm{T}}=\frac{2N+1}{2N}M_{\textrm{T}},
\end{align}
which leads to \eqref{eq:thres1linf}. When $\lambda\rightarrow 0$, we substitute $e^{-\lambda M_{\textrm{T}}}= 1-\lambda M_{\textrm{T}}$ into \eqref{eq:Threshold1}, which gives
\begin{align}
&\frac{1}{2}\alpha_{\textrm{th}}\rho  M_{\textrm{T}}-\frac{1-\lambda\alpha_{\textrm{th}}\rho  M_{\textrm{T}}}{2\lambda}+\frac{1}{2\lambda^2}=\frac{1}{2N} M_{\textrm{T}}-\frac{1-\lambda M_{\textrm{T}}}{2\lambda}+\frac{1}{2\lambda^2}.
\end{align}
Hence, we obtain \eqref{eq:thres1l0}.
\end{IEEEproof}
\end{Corollary}

From Corollary~\ref{Corollary:3}, we find the relationship between the number of UEs and the information ratio threshold, i.e., the information ratio threshold monotonically decreases when the number of UEs increases. This finding indicates that the DNP strategy is better than the BRNP strategy when the number of UEs is large. Moreover, we find that the information ratio threshold is high under the zero-waiting policy, but is low under the sporadic status generation rate. This finding indicates that the BRNP strategy is better for large $\lambda$ and the DNP strategy is better for small $\lambda$.

\subsection{Dynamic System}

In vehicular wireless networks, the number and locations of UEs may change from time to time. Thus, the information of each vehicle, such as its location, moving direction, and velocity, is required by other vehicles, implying that all vehicles may share a large amount of common information. In this subsection, we consider a dynamic system to characterize the features of a vehicular network \cite{Andrews2011}. In this system, we assume that the BS is located at the center and covers a circular annulus with inner radius $D_1$ and outer radius $D_2$, where $D_2>D_1\geq1$. In this area, we assume that the 
location of UEs follows a two-dimensional Poison point process (2D-PPP) with the intensity $\lambda_{UE}$. In this system, we assume that all UEs have the same common information and each UE has its own individual information. Here, we denote $L_{\mathrm{co}}$ as the number of bits of the common information and $L_{\mathrm{id}}$ as the number of bits of the individual information. Hence, with $N$ UEs in the system, the number of bits of the required status update for each UE is $L_h=L_{\mathrm{co}}+L_{\mathrm{id}}$ and the total number of bits of the status update is $L=L_{\mathrm{co}}+NL_{\mathrm{id}}$. We then denote $\beta$ as the ratio between the individual information and the common information, i.e., $\beta=\frac{L_{\mathrm{id}}}{L_{\mathrm{co}}}$. We note that the information ratio, $\alpha$, is given as $\alpha=\frac{N \beta+1}{N(\beta+1)}$ of the system with $N$ UEs. Since the information ration, $\alpha$, changes as the as number of UEs changes, we compare the AoI performance of two strategies based on the ratio between the individual information and the common information , $\beta$, in this system. 
We further assume that all UEs have the same path loss exponent, $\eta\geq 2$, and the received SNR at the BS is $\gamma=\gamma_0$. As per the path loss model, the received SNR of UE with the distance $d$ to the BS is $\gamma=\gamma_0d^{-\eta}$. Accordingly, we denote $C_0=\frac{1}{2}\log_2(1+\gamma_0)$ as the channel capacity with the received SNR $\gamma_0$, $C_{D_1}=\frac{1}{2}\log_2(1+\gamma_{D_1})$ as the channel capacity with the received SNR $\gamma_{D_1}=\gamma_0D_1^{-\eta}$, and $C_{D_2}=\frac{1}{2}\log_2(1+\gamma_{D_2})$ as the channel capacity with the received SNR $\gamma_{D_2}=\gamma_0D_2^{-\eta}$.
In the BRNP strategy, the BS transmits the status update with the coding rate $C_{D_2}$, i.e., $R=C_{D_2}$, which covers the entire area. In the DNP strategy, we assume that the BS detects the location of UEs and transmits their packets based on their location to minimize the average AoI of the system. Since the BS monitors the covered area in real time, the zero-waiting policy is adopted in this system.

Considering the dynamic system, 
we analyze and compare the expected average AoI achieved by the BRNP strategy and the DNP strategy in the following Theorem.
\begin{Theorem}\label{Theorem:Case2}
In the dynamic system, the expected average AoI in the BRNP strategy is approximated by
\begin{align}\label{eq:expaveAoIBr}
\E\left[\Delta_{\mathrm{BRNPZ}}\right]\approx \frac{3}{2C_{D_2}}\left(\left(1-e^{-\Lambda}\right)L_{\mathrm{co}}+\Lambda L_{\mathrm{id}}\right),
\end{align}
and the expected average AoI in the DNP strategy is approximated by
\begin{align}\label{eq:expaveAoID}
\E\left[\Delta_{\mathrm{DNPZ}}\right]\approx \left(\frac{1-e^{-\Lambda}}{2}+\Lambda\right)\left(\frac{L_{\mathrm{co}}+L_{\mathrm{id}}}{C_{\Lambda}}+M_{\textrm{L}}\right),
\end{align}
where $\Lambda=\lambda_{UE}\pi (D_2^2-D_1^2)$ and $C_{\Lambda} = \frac{(D_2^2-D_1^2)\eta}{   2^{2+\frac{4C_{0}}{\eta}}\ln{2}\left(\mathrm{Ei}(x_1)-\mathrm{Ei}(x_2)\right)}$. In $C_{\Lambda}$, $\mathrm{Ei}(x) = -\int_{-x}^{\infty}\frac{e^{-t}}{t}\mathrm{d}t$ is the exponential integral function, $x_1=-\frac{4}{\eta}C_{D_1}\ln{2}$, and $x_2 =-\frac{4}{\eta}C_{D_2}\ln{2}$.
\begin{IEEEproof}
See Appendix~\ref{Appendix:E}.
\end{IEEEproof}
\end{Theorem}

Based on Theorem~\ref{Theorem:Case2}, we compare the expected average AoI achieved by two schemes based on the ratio between the individual information and the common information, $\beta$, presented in Corollary~\ref{Corollary:case2x}.
\begin{Corollary}\label{Corollary:case2x}
In the dynamic system, the broadcast transmission scheme has a better AoI performance than the Unicast transmission scheme, if $\beta\leq\beta_{\textrm{th}}$, and the Unicast transmission scheme has a better AoI performance than the broadcast transmission scheme, if $\beta>\beta_{\textrm{th}}$. Specifically, $\beta_{\textrm{th}}$ is given by
\begin{align}
\beta_{\textrm{th}}=\frac{\frac{1-e^{-\Lambda}+2\Lambda}{C_{\Lambda}}- \frac{3(1-e^{-\Lambda})}{C_{D_2}}}{\frac{3\Lambda}{C_{D_2}}-\frac{1-e^{-\Lambda}+2\Lambda}{C_{\Lambda}}} +\frac{\left(2\Lambda+1-e^{-\Lambda}\right)M_{\textrm{L}}}{\left(\frac{3\Lambda}{C_{D_2}}-\frac{2\Lambda+1-e^{-\Lambda}}{C_{\Lambda}}\right)L_{\mathrm{co}}}.
\end{align}
\begin{IEEEproof}
The expected average AoI difference between the two schemes is approximated as
\begin{align}\label{eq:DeltaDf}
\Delta_{\textrm{Diff}}=&\frac{3}{2C_{D_2}}\left(\left(1-e^{-\Lambda}\right)L_{\mathrm{co}}+\Lambda\beta L_{\mathrm{co}}\right)-\left(\frac{1-e^{-\Lambda}}{2}+\Lambda\right)\left(\frac{(1+\beta) L_{\mathrm{co}}}{C_{\Lambda}}+M_{\textrm{L}}\right).
\end{align}
Based on \eqref{eq:DeltaDf}, we find that $\Delta_{\textrm{Diff}}\leq 0$ when $\beta\leq \beta_{\textrm{th}}$, and $\Delta_{\textrm{Diff}}> 0$ when $\beta> \beta_{\textrm{th}}$.
\end{IEEEproof}
\end{Corollary}

Corollary~\ref{Corollary:case2x} reveals that the BRNP strategy achieves the better AoI performance when the common information has a high proportion of the required status update of each UE, but the DNP strategy achieves the better AoI performance when the individual information has a high proportion of the required status update of each UE. Moreover, based on Corollary~\ref{Corollary:case2x}, we approximate the threshold $\beta_{\textrm{th}}$ as
\begin{align}
    \beta_{\textrm{th}}\approx \frac{2 C_{D_2}}{3 C_{\Lambda}-2C_{D_2}},
\end{align}
when $M_{\textrm{L}}$ is small and the average number of UEs in the area, denoted by $\Lambda$, is large. This indicates that the BRNP strategy achieves the better AoI performance if the BS covers a small area, where the difference in the received SNR among UEs is small, and the DNP strategy achieves the better AoI performance if the BS covers a large area, where the difference in the received SNR among UEs is large.

\section{Numerical Results}\label{Sec:Numerical}

In this section, we first present numerical results and evaluate the impact of various parameters, including the coding rate, the status update generation rate, and the number of UEs on the average AoI in the homogeneous case where all UEs have the same received SNR, i.e., $\gamma_n=\gamma_h$, and the same length of the required information, i.e., $L_n=L_h$, $n=1,2,\cdots,N$. We then examine how different transmission schemes affect the AoI performance in the considered remote control system and dynamic system.

\begin{figure}[!t]
\subfigure[]{
\begin{minipage}[b]{0.49\textwidth}
    \centering
    \includegraphics[width=0.98\columnwidth]{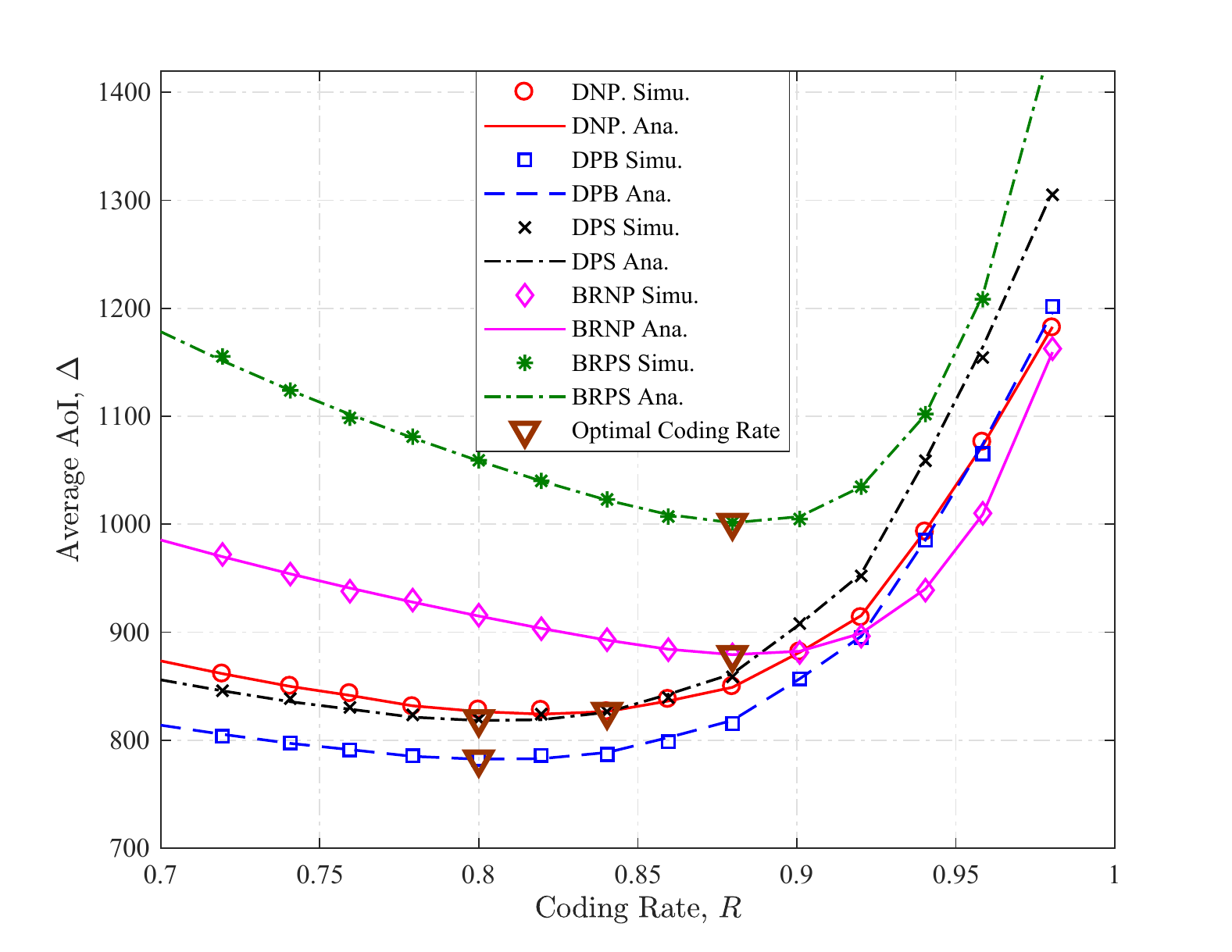}\label{fig:CodeR1}
\end{minipage}}
\subfigure[]{
\begin{minipage}[b]{0.49\textwidth}
    \centering
    \includegraphics[width=0.98\columnwidth]{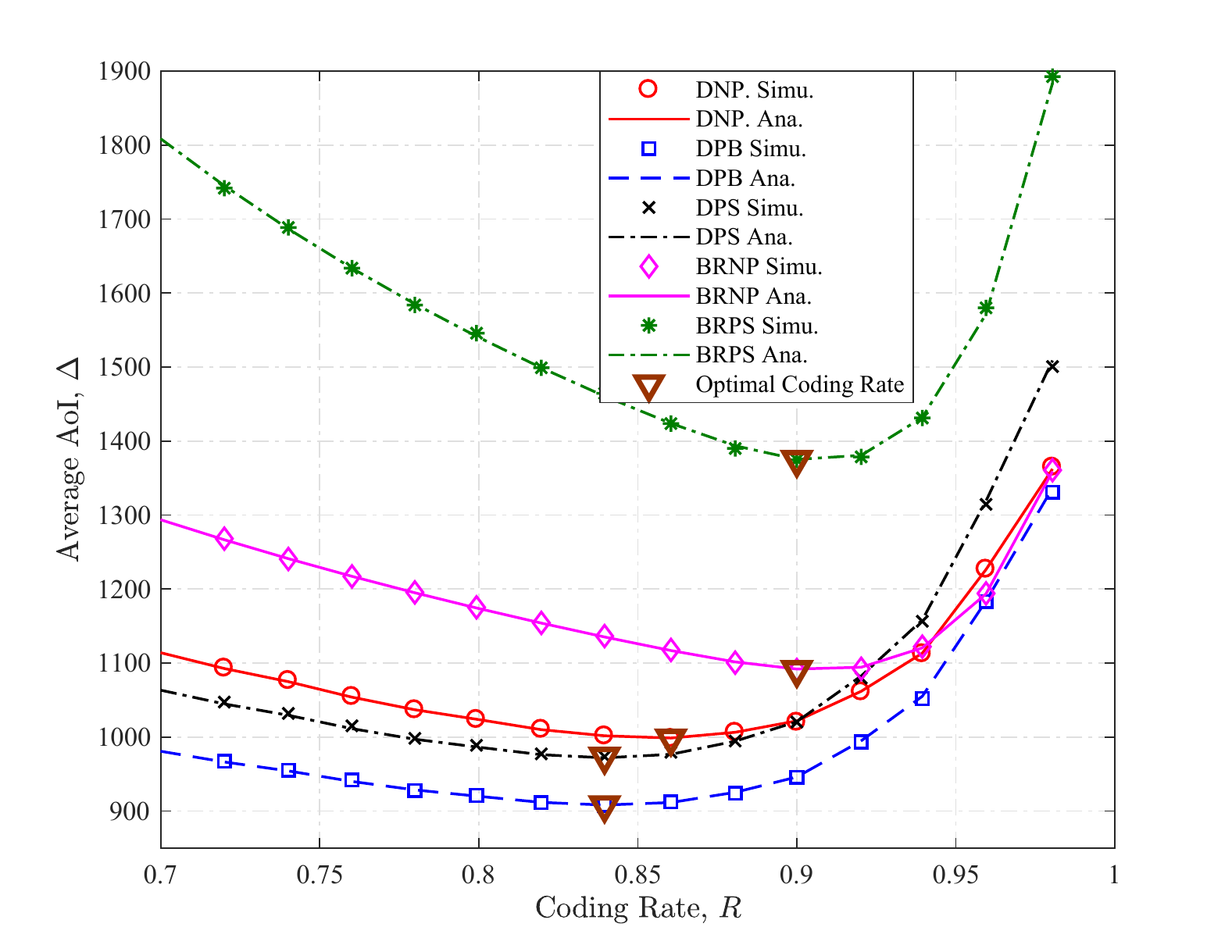}\label{fig:CodeR2}
\end{minipage}}
\caption{The average AoI versus the coding rate, $R$, with $N=3$, $\lambda = 0.002$, $\gamma_h=3$, $M_{\textrm{L}} = 10$, and $\alpha=1$, for (a) $L_h=100$ and (b) $L_h=150$.}\label{fig:CodeR}
\vspace{-1.5em}
\end{figure}

Fig.~\ref{fig:CodeR} plots the average AoI of the considered system versus the coding rate, $R$. We first observe that the analytical average AoI precisely matches the simulation results, which demonstrates the correctness of our analytical results in Theorem~\ref{Theorem:1}, Theorem~\ref{Theorem:2}, Theorem~\ref{Theorem:3}, and Corollary~\ref{Corollary:2}. We then observe that for all strategies, the average AoI first decreases and then increases when $R$ increases. This observation is due to the fact that the increase in $R$ has a two-fold effect on the average AoI. 
When $R$ is small, its increase leads to the shorter packet blocklength, which decreases the average AoI of the system. When $R$ exceeds a certain threshold, its increase leads to the significant increase in the block error rate, thereby degrading the AoI performance. We further observe that the optimal coding rate, which minimizes the average AoI, increases when the number of bits in a status update increases in all strategies. This is because that the increase in the number of bits in a status update leads to the increase in transmission time and the decrease in the block error rate. When the number of bits in a status update is low, the decrease in the block error rate dominantly and positively affects the average AoI. When the number of bits in a status update is high, the increase in transmission time dominates the average AoI, thereby degrading the AoI performance. Additionally, we observe that the BRNP strategy achieves a lower average AoI than the BRPS strategy, and the DPB strategy achieves the lowest average AoI compared with the DNP and DPS strategies. This observation reveals that the non-preemption strategy should be adopted in the broadcast transmission scheme and the preemption in buffer strategy should be adopted in the Unicast transmission scheme.

\begin{figure}[t]
\subfigure[]{
\begin{minipage}[b]{0.49\textwidth}
    \centering
    \includegraphics[width=0.98\columnwidth]{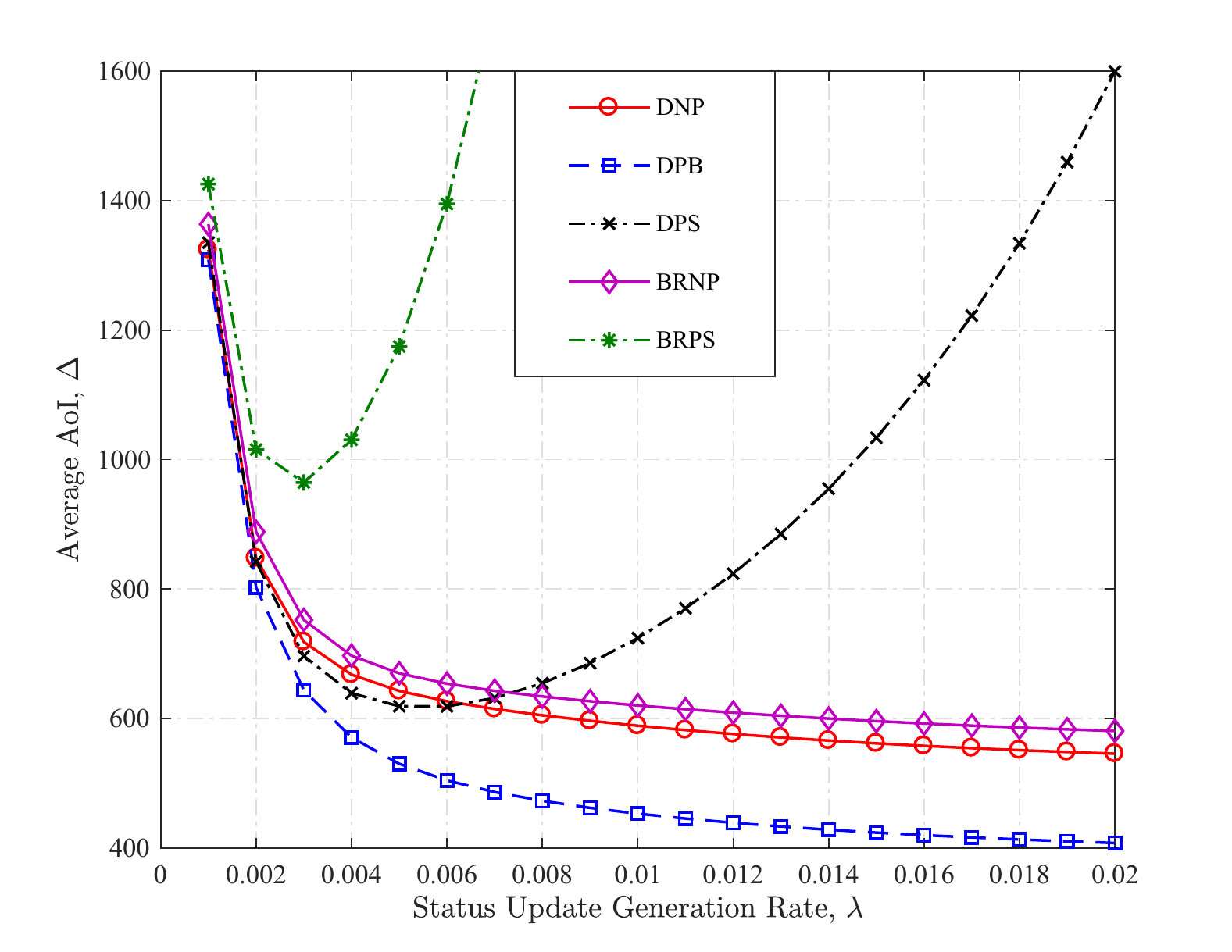}\label{fig:Lam1}
\end{minipage}}
\subfigure[]{
\begin{minipage}[b]{0.49\textwidth}
    \centering
    \includegraphics[width=0.98\columnwidth]{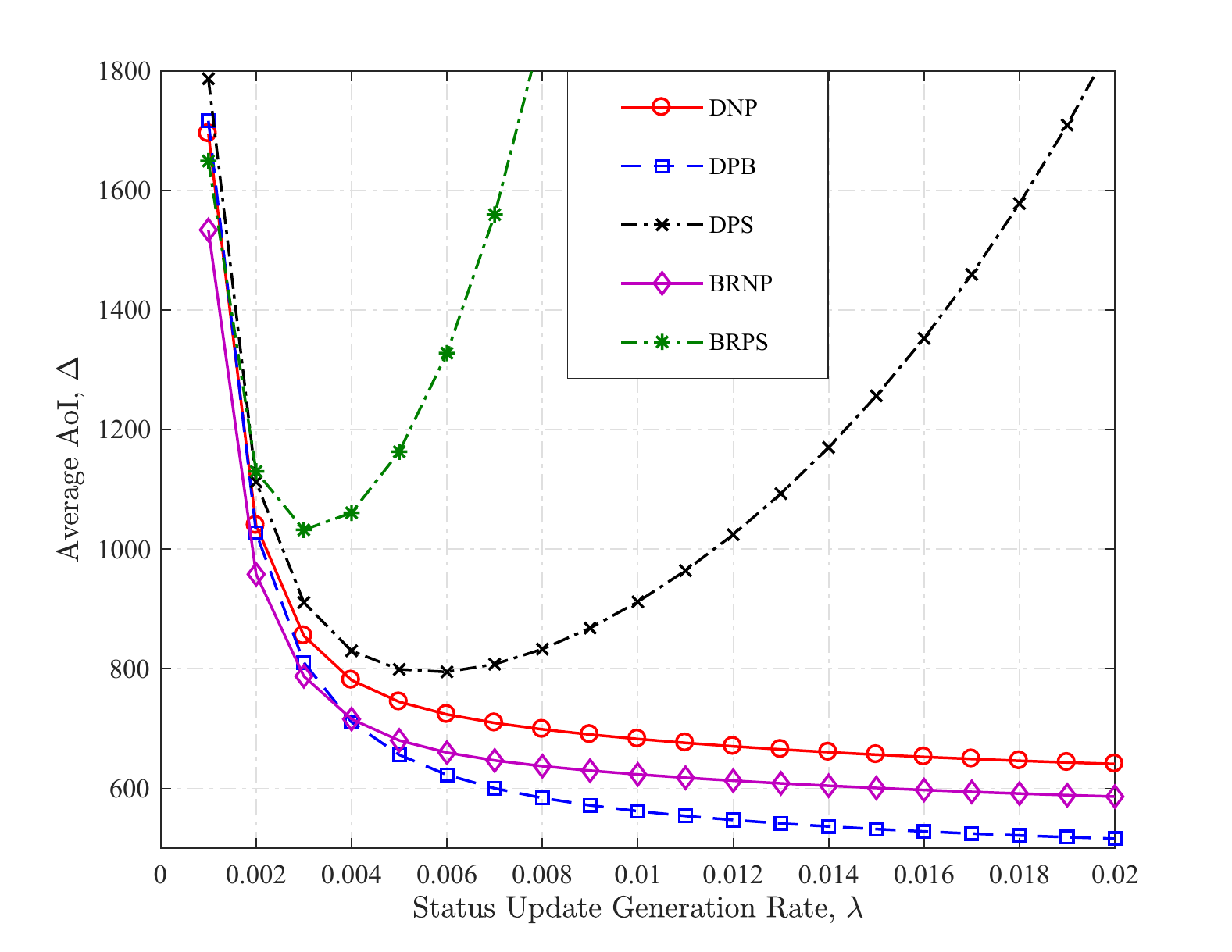}\label{fig:Lam2}
\end{minipage}}
\caption{The average AoI versus the packet generation rate, $\lambda$, with $N=3$, $L_h = 100$, $\gamma_h=3$, $M_{\textrm{L}} = 20$, and $\alpha=1$ for (a) $R=0.85$ and (b) $R=0.95$. }\label{fig:LAMBDA}
\vspace{-1.5em}
\end{figure}

Fig.~\ref{fig:LAMBDA} plots the average AoI of the considered system versus the status update generation rate, $\lambda$. We first observe that the average AoI decreases monotonically when $\lambda$ increases for the DNP, DPB, and BRNP strategies. This is because that the increase in $\lambda$ leads to the decrease in the waiting time after successful transmission, which decreases the average AoI of the system. We then observe that the average AoI first decreases and then increases when $\lambda$ increases for the DPS and BRPS strategies. This observation is due to the fact that the increase in $\lambda$ has a two-fold effect on the average AoI. When $\lambda$ is small, its increase reduces the waiting time after successful transmission, which improves the AoI performance. When $\lambda$ exceeds a certain threshold, its increase leads to the significant increase in the probability that a status update is preempted, thereby degrading the AoI performance.

\begin{figure}[t]
\subfigure[]{
\begin{minipage}[b]{0.49\textwidth}
    \centering
    \includegraphics[width=0.98\columnwidth]{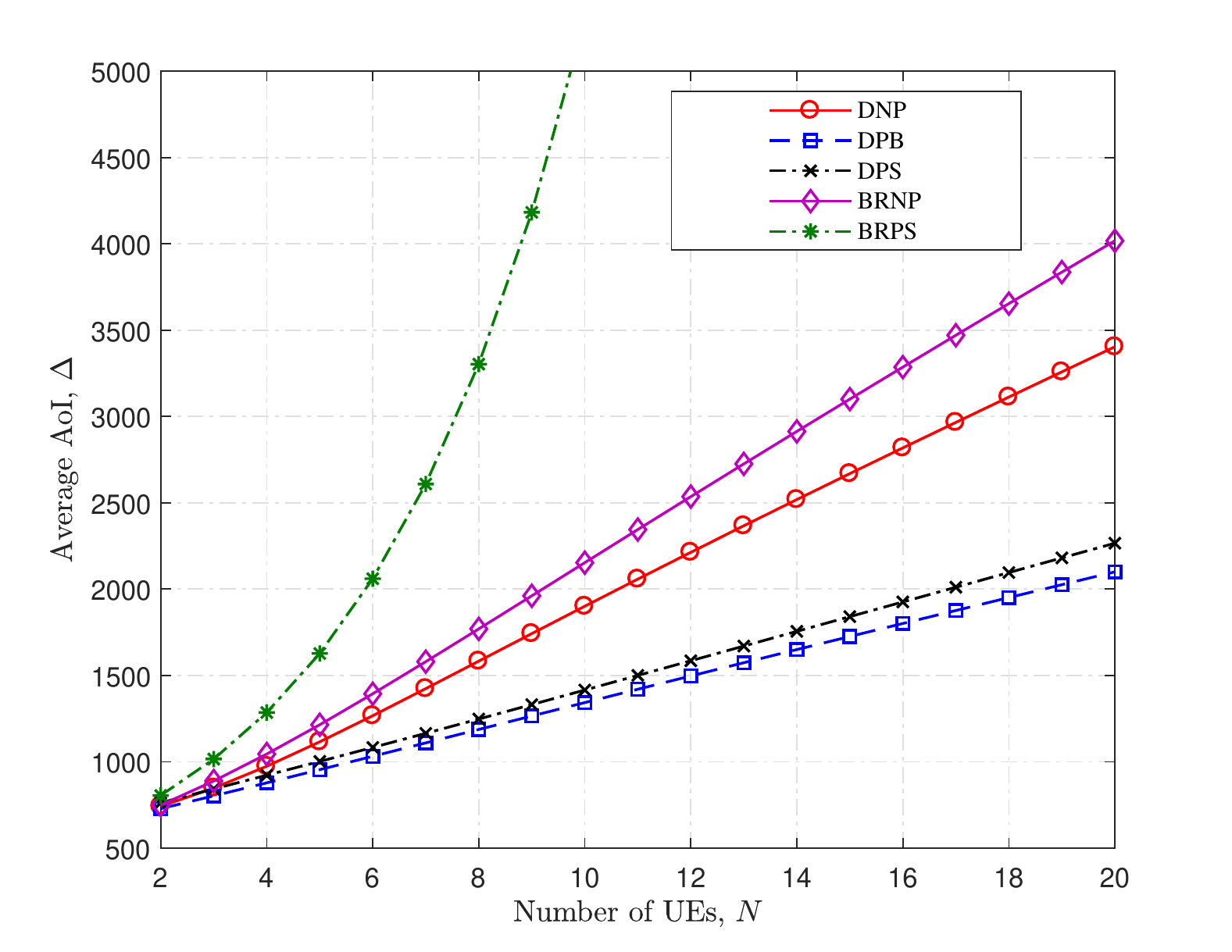}\label{fig:N1}
  
\end{minipage}
}
\subfigure[]{
\begin{minipage}[b]{0.49\textwidth}
    \centering
    \includegraphics[width=0.98\columnwidth]{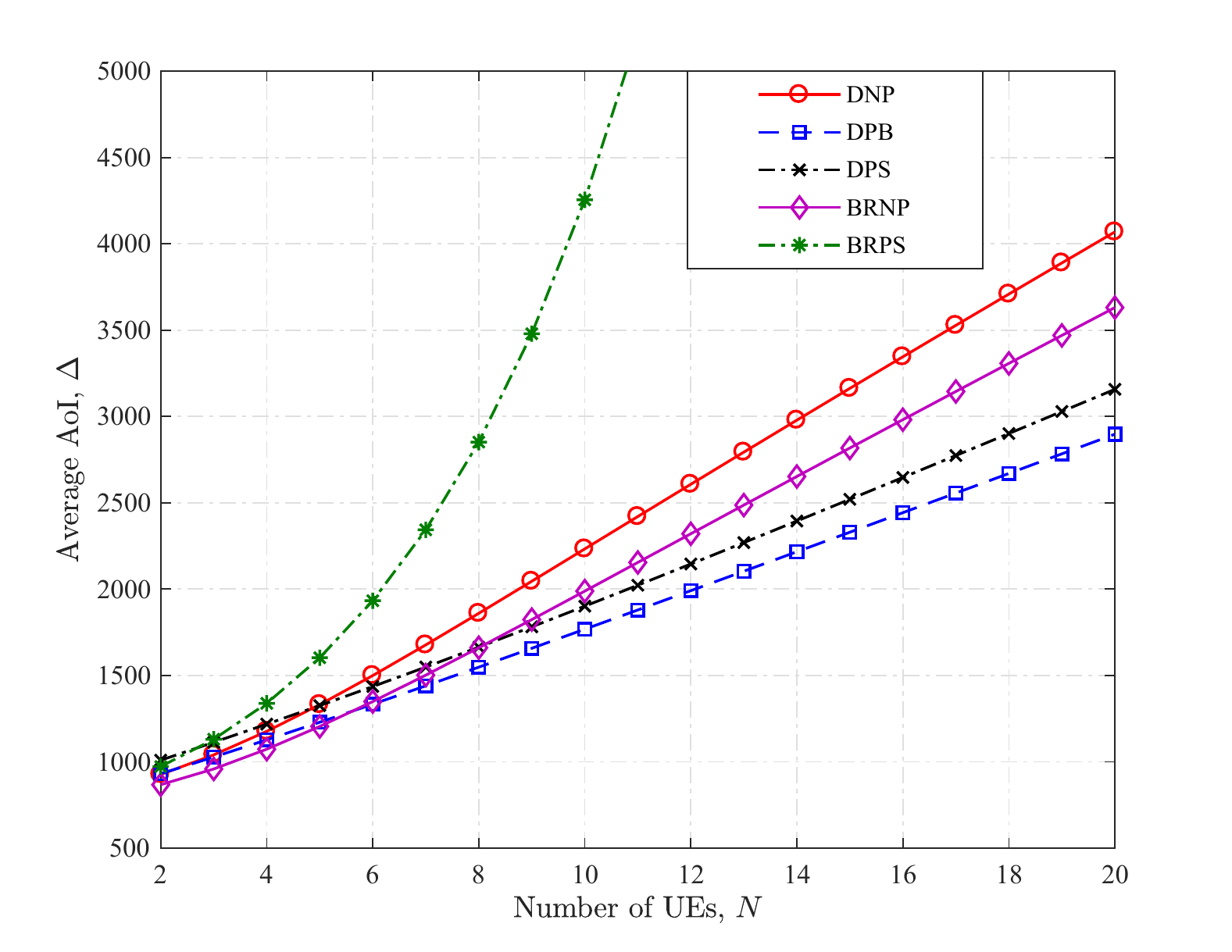}\label{fig:N2}
   
\end{minipage}
}
\caption{The average AoI versus the number of UEs, $N$, with $L_h = 100$, $\lambda=0.002$, $\gamma_h=3$, $M_{\textrm{L}} = 20$, and $\alpha=1$ for (a) $R=0.85$ and (b) $R=0.95$.}\label{fig:N}
\vspace{-1.5em}
\end{figure}

Fig.~\ref{fig:N} plots the average AoI of the considered system versus the number of sensors, $N$. We first observe that the average AoI achieved by all strategies increases monotonically when $N$ increases. This is because that the increase in $N$ results in the longer transmission time of status updates for all UEs, which increases the average AoI of the system. We then observe that when $N$ increases, the average AoI increases gently for the DPS strategy, but it increases dramatically for the BRPS strategy. In the BRPS strategy, the increase in $N$ leads to the increase in the probability that a status update is preempted, which increases the average AoI. Differently, in the DPS strategy, the increase in $N$ does not increase the probability that a status update is preempted for each UE, which leads to a gentle increase in the average AoI. This observation also implies that the preemption in serving is more suitable in the Unicast transmission scheme than in the broadcast transmission scheme. 
\begin{figure}[!t]
\centering
\includegraphics[width=0.95\columnwidth]{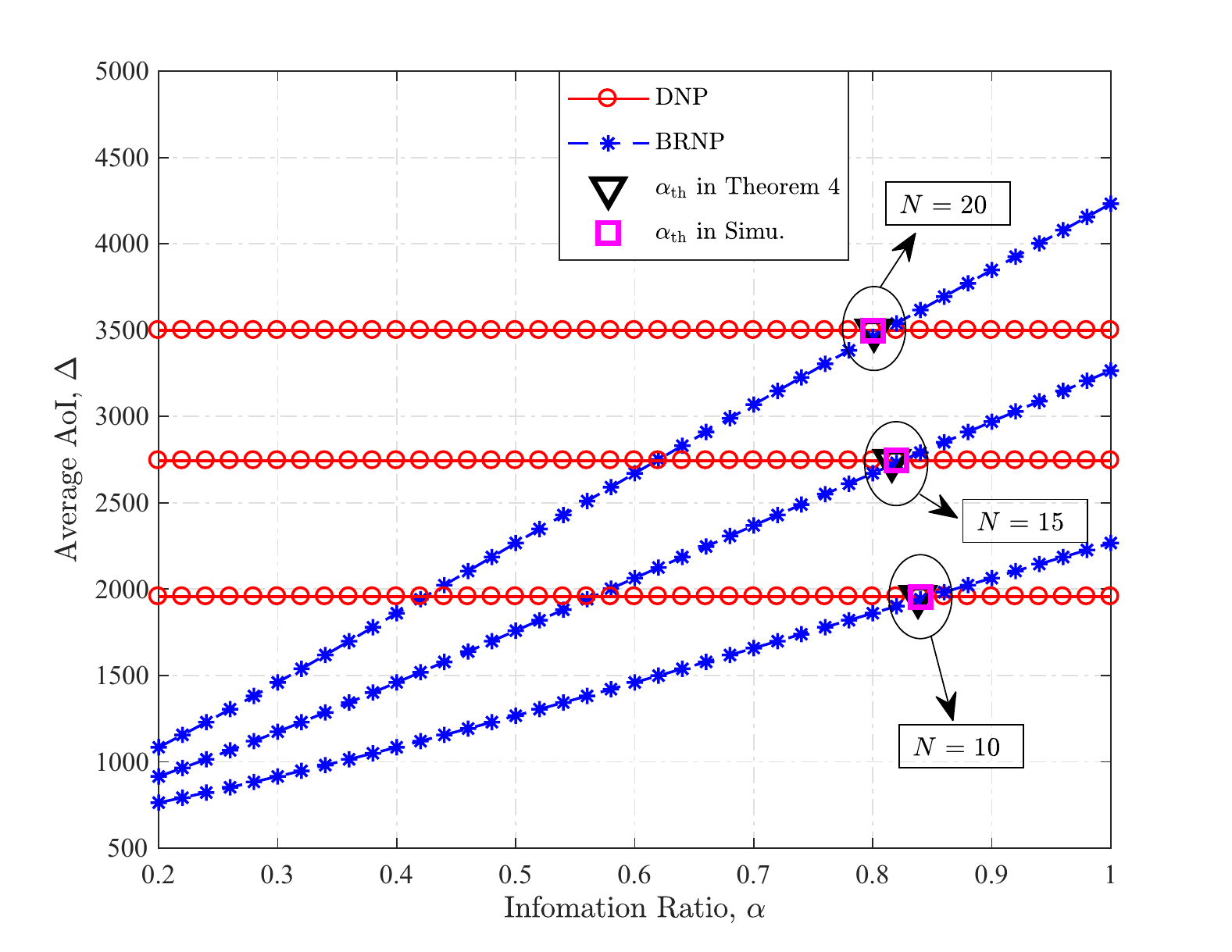}
\caption{The average AoI of the remote control system versus the information ratio, $\alpha$, with $L_h = 100$, $\lambda=0.002$, $\gamma_h=3$, $M_{\textrm{L}} = 20$, and $R=0.8$.}\label{fig:6ps}
\vspace{-1.5em}
\end{figure}

Fig.~\ref{fig:6ps} plots the average AoI of the remote control system versus the information ratio, $\alpha$. We first observe that the analytical $\alpha_{\mathrm{th}}$ precisely matches the simulation result, which demonstrates the correctness of our analytical result in Theorem~\ref{Lemma:1}. It indicates that we can adopt Theorem \ref{Lemma:1} to find the optimal transmission scheme to minimize the average AoI based on the information ratio. We then observe that the average AoI increases when $\alpha$ increases for the BRNP strategy. This is because that the increase in $\alpha$ leads to a larger number of bits in status updates for the transmission. Hence, the packet blocklength for transmission increases, which in turn increases the average AoI of the system. We further observe that $\alpha_{\mathrm{th}}$ decreases as the number of UEs increases. This observation is due to the fact that when $N$ is large, the DNP strategy significantly decreases the average waiting time of the status update for each UE, compared with the BRNP strategy, which decreases the average AoI. This observation implies that the BRNP strategy should be adopted when the number of UEs is small, while the DNP strategy should be adopted when the number of UEs is large.

\begin{figure}[!t]
\centering
\includegraphics[width=0.95\columnwidth]{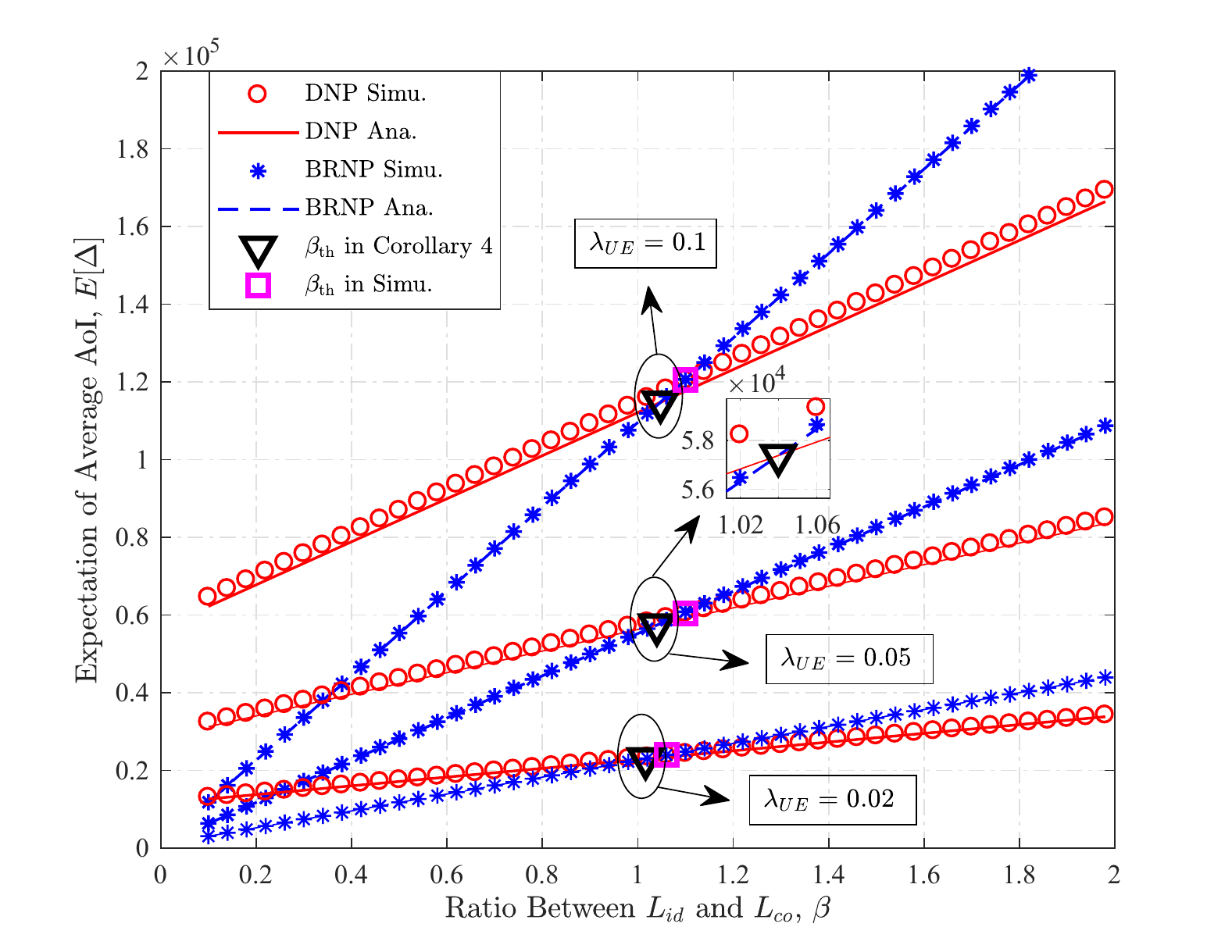}
\caption{The expected average AoI of the dynamic system versus the ratio between the individual information and the common information, $\beta$, with $L_{\mathrm{co}} = 1000$, $\gamma_{D_2}=10$, $M_{\textrm{L}} = 10$, $D_1=1$, $D_2=20$, and $\eta=2.2$.}\label{fig:7ps}
\vspace{-1.5em}
\end{figure}

Fig.~\ref{fig:7ps} plots the expected average AoI of the dynamic system versus the ratio between the individual information and the common information, $\beta$. We first observe that the approximation in Theorem \ref{Theorem:Case2} is close to the simulation result and there is a small gap between the approximation and the simulation result. This gap is due to the fact that the coding rate cannot achieve the channel capacity and the transmission error is inevitable in short packet communications. Due to the gap between the approximation and the simulation result, there is a gap between $\beta_{\mathrm{th}}$ in Corollary~\ref{Corollary:case2x} and the exact $\beta_{\mathrm{th}}$. We note that the gap between the expected average AoI in both schemes with $\beta_{\mathrm{th}}$ in Corollary~\ref{Corollary:case2x} is small. It indicates that we can adopt Corollary~\ref{Corollary:case2x} as a simple method to find the transmission scheme for minimizing the expected average AoI based on $\beta$. We further observe that the average AoI increases when $\beta$ increases for both strategies. Moreover, this increase in the BRNP strategy is faster than the DNP strategy. This is because that the increase in $\beta$ leads to the increase in the length of the status update in both strategies and this increase has a more pronounced impact on the average AoI for the BRNP strategy than the DNP strategy.  

\section{Conclusion}\label{Sec:Conclusion}
This paper analyzed the AoI performance achieved by two transmission schemes, i.e., the broadcast transmission scheme and the Unicast transmission scheme, of a multiuser system where a BS generates status updates and transmits them to multiple UEs. In both transmission schemes, we examined three packet management strategies, i.e., the non-preemption strategy, the preemption in buffer strategy, and the preemption in serving strategy. We first derived closed-form expressions for the average AoI achieved by these two transmission schemes with different packet management strategies. Such expressions allowed us to compare the AoI performance achieved by both transmission schemes in two systems, namely, the remote control system and the dynamic system. Aided by simulation results, we demonstrated the accuracy of our analysis. Moreover, we found that the Unicast transmission scheme is more appropriate for the system with a large number of UEs, while the broadcast transmission scheme is more appropriate for the system with a small number of UEs and high correlation among the information for UEs.

\begin{appendices}

\section{Proof of Theorem \ref{Theorem:1}}\label{Appendix:A}

To obtain $\Delta_{\mathrm{DNP},n}$, we need to derive $\E[T_j]$, $\E[Y_j]$, and $\E[Y_j^2]$ in \eqref{eq:newAoIevo}. We first derive $\E[T_j]$ by calculating it as
\begin{align}\label{eq:Tjcase1}
\E[T_j] = \E\left[S_j\right]+\E\left[W_{j}\right] = \sum_{k=1}^{n}M_k'+\E[W_{j}],
\end{align}
where $S_{j}$ is the serving time of $P_j$ from the time that $P_j$ is served by the BS to the time that $U_n$ receives $P_j$, and $W_j$ is the waiting time of $P_j$ in the buffer. We then derive $\E[W_{j}]$ as
\begin{align}\label{eq:Wjcase1}
\E[W_{j}]= \int_{0}^{M_{\textrm{T}}}x\lambda e^{-\lambda x}\mathrm{d}x=\frac{1-e^{-\lambda M_{\textrm{T}}}}{\lambda}-M_{\textrm{T}} e^{-\lambda M_{\textrm{T}}}.
\end{align}
By substituting \eqref{eq:Wjcase1} into \eqref{eq:Tjcase1}, we obtain $\E[T_j]$.

We then derive $\E[Y_j]$. Here, we denote $H_j$ as the number of status update transmitted to $U_n$ after $P_{j-1}$ is received by $U_n$ and before $P_j$ is received by $U_n$, and denote $P_{j,h}$ as the $h$-th status update transmitted to $U_n$ after $P_{j-1}$, where $P_{j,H_j}=P_j$. We note that
\begin{align}\label{eq:YjS2e}
\E[Y_j] = \E\left[\sum_{h=1}^{H_j} B_{j,h}\right] = \E[H_j]\E[B_{j,h}],
\end{align}
where $B_{j,h}$ is the time interval from the time that $P_{j,h-1}$ is transmitted to $U_n$ to the time that $P_{j,h}$ is transmitted to $U_n$, since $B_{j,h}$ is independent and identical distributed for $\forall h$. Here, we note that if $h = 1$, $P_{j,0}=P_{j-1}$. We find that $H_j$ follows a geometric distribution, where the probability mass function (PMF) of $H_j$ is given by
\begin{align}\label{eq:Kdistricase1}
\mathrm{Pr}(H_j=H) = (1-\epsilon_n)\epsilon_n^{H-1}.
\end{align}
According to \eqref{eq:Kdistricase1}, we obtain the expectation of $H_j$ and $H_j^2$, i.e., $\E[H_j]$ and $\E[H_{j}^2]$, are
\begin{align}\label{eq:EHj}
\E[H_j] = \frac{1}{1-\epsilon_n},
\end{align}
and 
\begin{align}\label{eq:EHj2}
\E[H_j^2] = \frac{1+\epsilon_n}{(1-\epsilon_n)^2},
\end{align}
respectively. In \eqref{eq:YjS2e}, $B_{j,h}$ is calculated by
\begin{align}\label{eq:Bjkcase1}
B_{j,h} = V_{j,h} + M_{\textrm{T}},
\end{align}
where $V_{j,h}$ is the time that the BS is in the idle state during $B_{j,h}$. Since $V_{j,k}$ follows an exponential distribution, we obtain $\E[V_{j,h}]$ as
\begin{align}\label{eq:Vjkcase1}
\E[V_{j,h}] = \int_{M_{\textrm{T}}}^{\infty}x\lambda e^{-\lambda x}\mathrm{d}x = \frac{1}{\lambda}e^{-\lambda M_{\textrm{T}}}.
\end{align}
By substituting \eqref{eq:EHj}, \eqref{eq:Bjkcase1}, and \eqref{eq:Vjkcase1} into \eqref{eq:YjS2e}, we obtain $\E[Y_j]$ as
\begin{align}\label{eq:Y1expcase1}
\E[Y_j] = \frac{1}{1-\epsilon_n}\left(M_{\textrm{T}}+\frac{e^{-\lambda M_{\textrm{T}}}}{\lambda}\right).
\end{align}

We next derive $\E\left[Y_j^2\right]$. Due to the independence and identical distribution of $B_{j,h_1}$ and $B_{j,h_2}$ when $h_1\neq h_2$, we calculate $\E\left[Y_j^2\right]$ as
\begin{align}\label{eq:Y2case1}
\E\left[Y_j^2\right]=&\E\left[\left(\sum_{h=1}^{H_j} B_{j,h}\right)^2\right]
=\E[H_j]\E\left[B_{j,h}^2\right]+\E\left[H_j^2-H_j\right]\E\left[B_{j,h}\right]^2.
\end{align}
Since $V_{j,k}$ follows an exponential distribution, we obtain $\E\left[V_{j,h}^2\right]$ as
\begin{align}\label{eq:Vjk2case2}
\E\left[V_{j,h}^2\right] = \int_{M_{\textrm{T}}}^{\infty}x^2\lambda e^{-\lambda x}\mathrm{d}x = \frac{2}{\lambda^2}e^{-\lambda M_{\textrm{T}}}.
\end{align}
By substituting \eqref{eq:EHj}, \eqref{eq:EHj2}, \eqref{eq:Bjkcase1}, and \eqref{eq:Vjk2case2} into \eqref{eq:Y2case1}, we obtain $\E\left[Y_j^2\right]$ as
\begin{align}\label{eq:Y2expcase1}
\E\left[Y_j^2\right] =\frac{\left(M_{\textrm{T}}+\frac{e^{-\lambda M_{\textrm{T}}}}{\lambda}\right)^2(1+\epsilon_n)}{(1-\epsilon_n)^2} + \frac{2e^{-\lambda M_{\textrm{T}}}-e^{-2\lambda M_{\textrm{T}}}}{(1-\epsilon_n)\lambda^2}.
\end{align}
By substituting \eqref{eq:Tjcase1}, \eqref{eq:Y1expcase1}, and \eqref{eq:Y2expcase1} into \eqref{eq:newAoIevo}, we obtain the final result given in \eqref{eq:expreAoIpar}, which completes this proof.

\section{Proof of Theorem \ref{Theorem:2}}\label{Appendix:B}

Based on \eqref{eq:newAoIevo}, we need to calculate $\E[T_j]$, $\E[Y_j]$, and $\E[Y_j^2]$ to obtain $\Delta_{\mathrm{DPB},n}$. We first derive $\E[T_j]$ by calculating it as
\begin{align}\label{eq:Tjcase2}
\E[T_j]=\E[S_j]+\E[W_{j}].
\end{align}
Here, we denote $A_{j,k}$ as the event that $P_{j}$ received by $U_n$ is generated when the BS is serving $U_{k_n}$ and $A_{j,0}$ as the event that $P_{j}$ received by $U_n$ is generated when the BS is in the idle state. Then, the probability of $A_{j,k}$ is obtained as
\begin{align}
\mathrm{Pr}\left(A_{j,k}\right)=\left\{
\begin{array}{ll}
\prod_{\kappa=k+1}^{N-1}p_k(1-p_{\kappa}),&\text{if }k\neq N_n \text{ and } k\neq0,\\
\prod_{\kappa=1}^{N}(1-p_{\kappa}),&\text{if }k=0,\\
\prod_{\kappa=1}^{N-1}p_{N}(1-p_{\kappa}),&\text{otherwise},
\end{array}
\right.
\end{align}
where $p_{k} = 1-e^{-\lambda M_{k_n}'}$ is the probability that at least one status update is generated when the BS is serving $U_{k_n}$. Hence, $\E[W_j]$ is calculated as
\begin{align}\label{eq:WjS2}
\E[W_j]=&\sum_{k=1}^{N}\frac{\mathrm{Pr}\left(A_{j,k}\right)}{p_k}\int_{0}^{M_{k_n}'}x\lambda e^{-\lambda x}\mathrm{d}x
=\frac{1-e^{-\lambda M_{\textrm{T}}}}{\lambda}-\sum_{k=1}^{N-1}M_{k_n}'e^{-\lambda\sum_{\kappa=k}^{N-1}M_{\kappa_n}'} - M_{N_n}'e^{-\lambda M_{\textrm{T}}}.
\end{align}
We denote $G_{j,k}$ as the event that the first packet of $P_j$ is transmitted to $U_{k_n}$, where the probability of $G_{j,k}$ is given by
\begin{align}\label{eq:Gjkcase2}
\mathrm{Pr}(G_{j,k})=\left\{
\begin{array}{ll}
\frac{\mathrm{Pr}(A_{j,k-1})}{1-e^{-\lambda M_{\textrm{T}}}},&\text{if}~k\neq 1,\\
\frac{\mathrm{Pr}(A_{j,N})}{1-e^{-\lambda M_{\textrm{T}}}},&\text{otherwise}.
\end{array}\right.
\end{align}
Based on \eqref{eq:Gjkcase2}, we calculate $\E[S_j]$ by
\begin{align}\label{eq:SjS2}
\E[S_j]=&\sum_{k=1}^{N}\mathrm{Pr}(G_{j,k})\sum_{\kappa=k}^{N}M_{\kappa_n}'
=\sum_{k=1}^{N}M_{k_n}'\frac{e^{-\lambda\sum_{\kappa=k}^{N-1}M_{\kappa_n}'}-e^{-\lambda M_{\textrm{T}}}}{1-e^{-\lambda M_{\textrm{T}}}}.
\end{align}
By substituting \eqref{eq:WjS2} and \eqref{eq:SjS2} into \eqref{eq:Tjcase2}, we obtain $\E[T_j]$ as
\begin{align}\label{eq:ETiS2}
\E[T_j]=&\frac{1\!-\!e^{-\lambda M_{\textrm{T}}}}{\lambda} \!-\!\sum_{k=1}^{N}M_{k_n}'\!\left(\!\frac{e^{-\lambda M_{\textrm{T}}}\left(1\!-\!e^{-\lambda\sum_{\kappa=k}^{N-1}M_{\kappa_n}'}\right)}{1-e^{-\lambda M_{\textrm{T}}}}\!\right)+M_{N_n}'(1-e^{-\lambda M_{\textrm{T}}}).
\end{align}

We then derive $\E[Y_j]$ based on \eqref{eq:YjS2e} and \eqref{eq:Bjkcase1}. We denote $Z_{j,k,h}$ as the event that the last packet of $P_{j,h-1}$ is for $U_{k_n}$ and the BS does not generate any status update during the service of $P_{j,h-1}$. Then, the probability of $Z_{j,k,h}$ is given by
\begin{align}\label{eq:PrGjhk}
\mathrm{Pr}\left(Z_{j,k,h}\right)=\frac{\left(1-e^{-\lambda M_{k_n}'}\right)e^{-\lambda M_{\textrm{T}}}}{1-e^{-\lambda M_{\textrm{T}}}}.
\end{align}
Hence, we calculate $\E\left[V_{j,h}\right]$ as
\begin{align}\label{eq:VjiS2}
\E\left[V_{j,h}\right]=& \frac{1}{\lambda}\sum_{k=1}^{N}\mathrm{Pr}\left(Z_{j,k,h}\right)
=\frac{e^{-\lambda M_{\textrm{T}}}}{\lambda(1-e^{-\lambda M_{\textrm{T}}})}\sum_{k=1}^{N}\left(1-e^{-\lambda M_{k_n}'}\right).
\end{align}
By substituting \eqref{eq:VjiS2} into \eqref{eq:Bjkcase1} and the result in \eqref{eq:YjS2e}, we obtain $\E\left[Y_j\right]$ as
\begin{align}\label{eq:EYjS2}
\E\left[Y_j\right]=\frac{1}{1-\epsilon_n}\left(M_{\textrm{T}}\!+\!\frac{e^{-\lambda M_{\textrm{T}}}}{\lambda(1-e^{-\lambda M_{\textrm{T}}})}\sum_{k=1}^{N}\left(1-e^{-\lambda M_{k_n}'}\right)\right).
\end{align}

We next derive $\E[Y_j^2]$. Based on the independence and identical distribution of $B_{j,h_1}$ and $B_{j,h_2}$ when $h_1\neq h_2$, we calculate $\E\left[Y_j^2\right]$ from \eqref{eq:Y2case1} as
\begin{align}\label{eq:EYj2S2}
\E\left[Y_j^2\right] 
=&\frac{1}{1-\epsilon}\left(M_{\textrm{T}}^2+2M_{\textrm{T}}\E\left[V_{j,h}\right]
+\E\left[V_{j,h}^2\right]\right)
+\frac{2\epsilon}{\left(1-\epsilon\right)^2}\left(M_{\textrm{T}}+\xi\right)^2\notag\\
=&\frac{1+\epsilon}{\left(1-\epsilon\right)^2}\left(M_{\textrm{T}}+\xi
\right)^2+ \frac{2\xi-\lambda \xi^2}{\lambda\left(1-\epsilon\right)},
\end{align}
where $\E[V_{j,h}^2] = \frac{2}{\lambda^2}\sum_{k=1}^{N}\mathrm{Pr}\left(Z_{j,k,h}\right)$ and $\E[V_{j,h}]$ is given in \eqref{eq:VjiS2}. By substituting \eqref{eq:ETiS2}, \eqref{eq:EYjS2}, and \eqref{eq:EYj2S2} into \eqref{eq:newAoIevo}, we obtain the final result given in \eqref{eq:expreAoIS2}, which completes this proof.

\section{Proof of Theorem \ref{Theorem:3}}\label{Appendix:C}

Based on \eqref{eq:newAoIevo}, we need to calculate $\E[T_j]$, $\E[Y_j]$, and $\E[Y_j^2]$ to obtain $\Delta_{\mathrm{DPS},n}$. We first derive $\E[T_j]$. In the DPS strategy, the waiting time of $P_j$ is $0$, i.e., $W_j=0$, since preemption is considered. Thus, we calculate $\E[T_j]$ as
\begin{align}
\E[T_j] = \E[S_j].
\end{align}
We note that in this strategy, the probability of $A_{j,k}$ and $G_{j,k}$ are given by
\begin{align}
\mathrm{Pr}\left(A_{j,k}\right) = \left\{
\begin{array}{ll}
\prod_{\kappa=k}^{N}p_k(1-p_{\kappa}),&\text{if}~k\neq 0,\\
\prod_{\kappa=1}^{N}(1-p_{\kappa}),&\text{otherwise},
\end{array}\right.
\end{align}
and
\begin{align}\label{eq:Gjkcase3}
\mathrm{Pr}\left(G_{j,k}\right)=\frac{1}{1-e^{-\lambda M_{\textrm{T}}}}\mathrm{Pr}\left(A_{j,k}\right),
\end{align}
respectively. Based on \eqref{eq:Gjkcase3}, we calculate $\E[T_j]$ as
\begin{align}\label{eq:SjS3}
\E[T_j]=&\E[S_j]=\sum_{k=1}^{N}\mathrm{Pr}\left(G_{j,k}\right)
\sum_{\kappa=k}^{N}M_{\kappa_n}'
=\sum_{k=1}^{N}M_{k_n}'\frac{e^{-\lambda\sum_{\kappa=k+1}^{N}M_{\kappa_n}'}-e^{-\lambda M_{\textrm{T}}}}{1-e^{-\lambda M_{\textrm{T}}}}.
\end{align}

We then derive $\E[Y_j]$ based on \eqref{eq:YjS2e}. We note that
\begin{align}\label{eq:BjiS3}
B_{j,h}=V_{j,h}+\sum_{k=1}^{N}S_{j,k,h},
\end{align}
where $S_{j,k,h}$ is the serving time of the $U_{k_n}$'s packet during $B_{j,h}$. We then note that the probability of $Z_{j,k,h}$ is given by
\begin{align}\label{eq:PrGjhkcase3}
\mathrm{Pr}\left(Z_{j,k,h}\right)=\frac{(1-e^{-\lambda M_{k+1_n}'})e^{-\lambda M_{\textrm{T}}}}{1-e^{-\lambda M_{\textrm{T}}}},
\end{align}
for the DPS strategy. Based on \eqref{eq:PrGjhkcase3}, we calculate $\E\left[V_{j,h}\right]$ as
\begin{align}\label{eq:EVjhcase3}
\E\left[V_{j,h}\right]=\frac{1}{\lambda}\sum_{k=1}^{N}\mathrm{Pr}\left(Z_{j,k,h}\right)
=\frac{e^{-\lambda M_{\textrm{T}}}}{\lambda(1-e^{-\lambda M_{\textrm{T}}})}\sum_{k=1}^{N}\frac{p_k}{1-p_k}.
\end{align}
For $S_{j,k,h}$, it is calculated as
\begin{align}\label{eq:Sjkhcase3}
S_{j,k,h}=\sum_{\theta=0}^{\Theta_k}S_{j,k,h,\theta}+M_{k_n}',
\end{align}
where $\Theta_k$ is the times of preemption of status update for $U_{k_{n}}$ and $S_{j,k,h,\theta}$ is the serving time of the $\theta$-th preempted status update. We note that $\Theta_k$ follows a geometric distribution, i.e., $\mathrm{Pr}\left(\Theta_k=\Theta\right) = (1-p_k)p_k^\Theta$, and the average serving time of the $\theta$-th preempted status update is calculated as
\begin{align}\label{eq:Sjkhtheta}
\E\left[S_{j,k,h,\theta}\right]=\frac{1}{p_k}\int_0^{M_{k_n}'}\lambda t e^{-\lambda t}\mathrm{d}t=\frac{1}{\lambda}-\varphi,
\end{align}
where $\varphi=\frac{\left(1-p_k\right)M_{k_n}'}{p_k}$. Combining \eqref{eq:Sjkhcase3} with \eqref{eq:Sjkhtheta}, we calculate $\E\left[S_{j,k,h}\right]$ by
\begin{align}\label{eq:Sjhkcase3}
\E\left[S_{j,k,h}\right]=&\E\left[\Theta_k]\E[S_{j,k,h,\theta}\right]+M_{k_n}'
=\frac{p_k}{1-p_k}\left(\frac{1}{\lambda}-\varphi\right)+M_{k_n}'=\frac{p_k}{\lambda(1-p_k)}.
\end{align}
By substituting \eqref{eq:EVjhcase3} and \eqref{eq:Sjhkcase3} into \eqref{eq:BjiS3}, we obtain $\E[B_{j,h}]$ as
\begin{align}\label{eq:Bjhcase3}
\E[B_{j,h}] 
=\frac{1}{\lambda(1-e^{-\lambda M_{\textrm{T}}})}\sum_{k=1}^{N}\frac{p_k}{1-p_k}.
\end{align}
Then, by substituting \eqref{eq:Bjhcase3} into \eqref{eq:YjS2e}, we obtain $\E[Y_j]$ as
\begin{align}\label{eq:Yjcase3}
\E[Y_j]=\frac{1}{\lambda(1-\epsilon_n)(1-e^{-\lambda M_{\textrm{T}}})}\sum_{k=1}^{N}\frac{p_k}{1-p_k}.
\end{align}

We next calculate $\E\left[Y_j^2\right]$ based on \eqref{eq:Y2case1}. According to \eqref{eq:BjiS3}, we calculate $\E\left[B_{j,h}^2\right]$ in \eqref{eq:Y2case1} as
\begin{align}\label{eq:EBjhcase3}
\E\left[B_{j,h}^2\right] 
=&\E\left[V_{j,h}^2\right]+2\E\left[V_{j,h}\left(\sum_{k=1}^{N}S_{j,k,h}\right)\right]+\E\left[\left(\sum_{k=1}^{N}S_{j,k,h}\right)^2\right].
\end{align}
We note that $V_{j,h}$ is not independent from $S_{j,k,h}$ in \eqref{eq:EBjhcase3}. If the BS is in the idle state before transmitting the packet of $U_{k_n}$, the packet of $U_{\kappa_n}$, $\kappa\in\{1,2,\cdots,k\}$, is not preempted during $B_{j,h}$. Hence, we obtain
\begin{align}\label{eq:VJSJS3}
\E\left[V_{j,h}\left(\sum_{k=1}^{N}S_{j,k,h}\right)\right]=&\frac{e^{-\lambda M_{\textrm{T}}'}}{\lambda(1-e^{-\lambda M_{\textrm{T}}'})}\sum_{k=1}^N\frac{p_k}{1-p_k}\left(\sum_{\kappa=k}^{N}
\frac{p_{\kappa}}{\lambda(1-p_{\kappa})}\right).
\end{align}
We then calculate $\E\left[V_{j,h}^2\right]$ and $\E\left[\left(\sum_{k=1}^{N}S_{j,k,h}\right)^2\right]$ in \eqref{eq:EBjhcase3} as
\begin{align}\label{eq:VJ2S3}
\E\left[V_{j,h}^2\right] = \frac{2e^{-\lambda M_{\textrm{T}}}}{\lambda^2(1-e^{-\lambda M_{\textrm{T}}})}\sum_{k=1}^{N}\frac{p_k}{1-p_k},
\end{align}
and
\begin{align}\label{eq:Sj2S3}
\E\left[\left(\sum_{k=1}^{N}S_{j,k,h}\right)^2\right]
&=\sum_{k=1}^{N}\frac{p_k}{1-p_k}\left(\frac{2}{\lambda^2}-\frac{2\varphi}{\lambda}
-\varphi M_{k_n}'\right)\notag\\
&+\sum_{k=1}^{N}\sum_{\kappa\neq k}\frac{p_k}{1-p_k}\left(\frac{1}{\lambda}-\varphi\right)
\frac{p_{\kappa}}{1-p_{\kappa}}\left(\frac{1}{\lambda}
-\frac{M_{\kappa_n}'(1-p_\kappa)}{p_\kappa}\right)\notag\\
&+\sum_{k=1}^{N}\frac{2p_k^2}{(1-p_k)^2}\left(\frac{1}{\lambda}-\varphi\right)^2
+2M_{\textrm{T}}\sum_{k=1}^{N}\left(\frac{p_k}{1-p_k}
\left(\frac{1}{\lambda}-\varphi\right)\right)+M_{\textrm{T}}^2,
\end{align}
respectively. By substituting \eqref{eq:VJSJS3}, \eqref{eq:VJ2S3}, and \eqref{eq:Sj2S3} into \eqref{eq:EBjhcase3}, and combining \eqref{eq:EBjhcase3} with \eqref{eq:Y2case1}, we obtain $\E\left[Y_{j}^2\right]$. By substituting \eqref{eq:SjS3}, \eqref{eq:Yjcase3}, and $\E\left[Y_{j}^2\right]$ into \eqref{eq:newAoIevo}, we obtain the final result in \eqref{eq:expreAoIS3}, which completes the proof.

\section{Proof of Theorem 4}\label{Appendix:D}

In the remote control system, we find that the block error rate $\epsilon_n$ is negligible, as a low coding rate is assumed. With the information ratio, $\alpha$, and the same coding rate, $R$, we obtain the transmitted packet length, $M$, in the BRNP strategy as
\begin{align}
M = \frac{L}{R} = \frac{\alpha NL_h}{R}=\alpha \rho M_{\textrm{T}}.
\end{align}
Hence, the average AoI in the BRNP strategy is calculated as
\begin{align}
\Delta_{\mathrm{BRNP}}&=\frac{1}{2}\left(M+\frac{1}{\lambda}e^{-\lambda M}\right)+\frac{2e^{-\lambda M}-e^{-2\lambda M}}{2\lambda^2(M+\frac{1}{\lambda}e^{-\lambda M})}
+\left(\frac{1}{\lambda}+M\right)\left(1-e^{-\lambda M}\right)\notag\\
=&\frac{1}{2}\left(\alpha\rho M_{\textrm{T}}\!+\!\frac{1}{\lambda}e^{-\lambda\alpha\rho  M_{\textrm{T}}}\right)\!+\!\frac{2e^{-\lambda\alpha\rho  M_{\textrm{T}}}-e^{-2\lambda\alpha\rho  M_{\textrm{T}}}}{2\lambda^2(\alpha\rho M_{\textrm{T}}+\frac{1}{\lambda}e^{-\lambda\alpha\rho  M_{\textrm{T}}})}
+\left(\frac{1}{\lambda}+\alpha\rho M_{\textrm{T}}\right)\left(1-e^{-\lambda \alpha\rho M_{\textrm{T}}}\right).
\end{align}
Since all UEs require the same length of the status update, $L_h$, and have the same coding rate $R$, the system average AoI in the DNP strategy is calculated as
\begin{align}
\Delta_{\mathrm{DNP}}=&\frac{1}{2}\left(M_{\textrm{T}}+\frac{1}{\lambda}e^{-\lambda M_{\textrm{T}}}\right)+\frac{2e^{-\lambda M_{\textrm{T}}}-e^{-2\lambda M_{\textrm{T}}}}{2\lambda^2(M_{\textrm{T}}+\frac{1}{\lambda}e^{-\lambda M_{\textrm{T}}})}
-\frac{N-1}{2N}M_{\textrm{T}}\!+\!\left(\frac{1}{\lambda}+M_{\textrm{T}}\right)(1-e^{-\lambda M_{\textrm{T}}}).
\end{align}
By calculating the average AoI difference between the BRNP strategy and the DNP strategy, we obtain the final result in \eqref{eq:Threshold1}.

\section{Proof of Theorem~\ref{Theorem:Case2}}\label{Appendix:E}

In the dynamic system, we denote $N_{\Lambda}$ as the number of UEs, which follows a Poisson distribution with the PMF given by
\begin{align}
\mathrm{Pr}(N_{\Lambda} = N) = \frac{\Lambda^{N}e^{-\Lambda}}{N!}.
\end{align}

We first analyze the expected average AoI in the BRNP strategy. We assume that the coding rate of transmission is close to the channel capacity and the block error rate is negligible. Then, the average AoI of the system with $N$ UEs, where $N\neq0$, is approximated as
\begin{align}
\Delta_{\mathrm{BRNPZ}}^{(N)} \approx \frac{3(L_{\mathrm{co}}+N L_{\mathrm{id}})}{2C_{D_2}}.
\end{align}
By averaging the average AoI with respect to the number of UEs, the expected average AoI is calculated as
\begin{align}
\E[\Delta_{\mathrm{BRNPZ}}] =&\sum_{N=1}^{\infty}\mathrm{Pr}(N_{\Lambda} = N)\Delta_{\mathrm{BRNPZ}}^{(N)}
\approx\sum_{N=1}^{\infty}\frac{3\Lambda^N e^{-\Lambda}(L_{\mathrm{co}}+N L_{\mathrm{id}})}{2N!C_{D_2}},
\end{align}
which results in \eqref{eq:expaveAoIBr}.

We then analyze the expected average AoI in the DNP strategy. Here, we approximate the average AoI by considering the coding rate to be close to the channel capacity of each UE and neglecting the block error rate. Hence, we approximate the average AoI of the system with $N$ UEs, where $N\neq0$, as
\begin{align}\label{eq:DNPNcase2}
\Delta_{\mathrm{DNPZ}}^{(N)}\approx&\frac{3}{2}\sum_{n=1}^N
\left(\E\left[\frac{L_{\mathrm{co}}+L_{\mathrm{id}}}{C_n}\right]+M_{\textrm{L}}\right)-\frac{1}{N}\sum_{n=1}^N\sum_{k=n+1}^N\left(\E\left[\frac{L_{\mathrm{co}}
+L_{\mathrm{id}}}{C_n}\right]+M_{\textrm{L}}\right),
\end{align}
where $C_n$ is the channel capacity of $U_n$. Due to the fact that the location of UEs follows the same Poisson process, we simplify \eqref{eq:DNPNcase2} as
\begin{align}\label{eq:DNPNcase2x}
\Delta_{\mathrm{DNPZ}}^{(N)} \approx \frac{2N+1}{2}\left(\left(L_{\mathrm{co}}+L_{\mathrm{id}}\right)\E\left[\frac{1}{C_n}\right]+M_{\textrm{L}}\right).
\end{align}
To obtain $\Delta_{\mathrm{DNPZ}}^{(N)}$, we need to calculate $\E\left[1/C_{n}\right]$. Since the distance from each UE to the BS follows a 2D Poisson process, the probability density function (PDF) of the distance between the BS and each UE is given by
\begin{align}\label{eq:pdfdcase2}
f(d) = \frac{2d}{D_2^2-D_1^2}.
\end{align}
With high received SNR, the channel capacity is approximated as
\begin{align}\label{eq:channelCaprx}
C_n = \frac{1}{2}\log_2\left(1+\gamma_n\right)\approx\frac{1}{2}\log_2\left(\gamma_n\right).
\end{align}
Based on \eqref{eq:pdfdcase2} and \eqref{eq:channelCaprx}, we calculate $\E\left[1/C_{n}\right]$ as
\begin{align}\label{eq:Cnincase2}
\E\left[\frac{1}{C_n}\right] \approx&\int_{D_1}^{D_2}\frac{1}{C_0-\frac{\eta}{2}\log_2 d}f(d)\mathrm{d}d
=\frac{2^{2+\frac{4C_{0}}{\eta}}\ln{2}}{\left(D_2^2-D_1^2\right)\eta}
\left(\mathrm{Ei}\left(x_1\right)-\mathrm{Ei}\left(x_2\right)\right)=\frac{1}{C_{\Lambda}}.
\end{align}
By substituting \eqref{eq:Cnincase2} into \eqref{eq:DNPNcase2x}, we obtain the average AoI of the system with $N$ UEs in the DNP strategy as
\begin{align}
\Delta_{\mathrm{DNPZ}}^{(N)}\approx\frac{2N+1}{2}\left(\frac{L_{\mathrm{co}}+L_{\mathrm{id}}}
{C_{\Lambda}}+M_{\textrm{L}}\right).
\end{align}
By averaging the average AoI with respect to the number of UEs, the expected average AoI is calculated as
\begin{align}
\E[\Delta_{\mathrm{DNPZ}}]=&\sum_{N=1}^{\infty}\mathrm{Pr}(N_{\Lambda} = N)\Delta_{\mathrm{DNPZ}}^{(N)}
\approx\sum_{N=1}^{\infty}\frac{\Lambda^N e^{-\Lambda}\left(2N+1\right)}{2N!}\left(\frac{L_{\mathrm{co}}+L_{\mathrm{id}}}{C_n}+M_{\textrm{L}}\right),
\end{align}
which results in \eqref{eq:expaveAoID}.

\end{appendices}

\bibliographystyle{IEEEtran} 
\bibliography{bibli}

\end{document}